\documentclass[aps,prx,twocolumn]{revtex4-1}
\usepackage{dcolumn}
\usepackage{graphicx}
\usepackage{bm}
\usepackage{color}
\usepackage{xcolor}
\usepackage{amsmath}
\usepackage{amstext}
\usepackage{amssymb}
\usepackage{booktabs}
\usepackage{array}
\usepackage[colorlinks=true,linkcolor=blue,citecolor=blue,urlcolor=blue]{hyperref}
\usepackage{ulem}
\usepackage{verbatim} 

\newcommand{\suppinfo}{Supplemental Material~\cite{supp-info}}

\renewcommand{\emph}[1]{\textit{#1}}

\newcommand{\editor}[2]{%
  \expandafter\newcommand\csname #1note\endcsname[1]{%
    \textcolor{#2}{(\textbf{#1:} ##1)}}%
  \expandafter\newcommand\csname #1\endcsname[1]{%
    \textcolor{#2}{##1}}%
  \expandafter\newcommand\csname #1cancel\endcsname[1]{%
    \textcolor{#2}{\sout{##1}}}%
  \expandafter\newcommand\csname #1change\endcsname[2]{%
    \textcolor{#2}{\sout{##1} ##2}}%
  \newenvironment{#1text}{\color{#2}}{\color{black}}
}
\definecolor{Blu}{rgb}{0.00,0.00,1.00}
\definecolor{Red}{rgb}{1.00,0.00,0.00}
\definecolor{Cyan}{rgb}{0.00,0.50,0.50}
\definecolor{Green}{rgb}{0.00,0.70,0.00}
\definecolor{BluBondi}{rgb}{0.00,0.58,0.71}
\editor{DV}{blue}
\editor{AF}{Red}
\editor{CC}{Cyan}
\editor{DAL}{Green}
\editor{EM}{orange}
\editor{TC}{violet}
\editor{CH}{BluBondi}   

\begin{document}

\title{Frequency dependence in GW made simple using a multi-pole approximation}

\author{Dario A. Leon$^{1,2}$, Claudia Cardoso$^{2}$, Tommaso Chiarotti$^{3}$, Daniele Varsano$^{2}$, Elisa Molinari$^{1,2}$, and Andrea Ferretti$^2$}
\affiliation{$^1$FIM Department, University of Modena \& Reggio Emilia, Via Campi 213/a, Modena (Italy)}
\affiliation{$^2$S3 Centre, Istituto Nanoscienze, CNR, Via Campi 213/a, Modena (Italy)}
\affiliation{$^3$Theory and Simulation of Materials (THEOS), Ecole Polytechnique F\'ed\'erale de Lausanne (EPFL), CH-1015 Lausanne (Switzerland)}

\begin{abstract}
In the $GW$ approximation, the screened interaction $W$ is a non-local and dynamical potential that usually has a complex frequency dependence. 
A full description of such dependence is possible but often computationally demanding. 
For this reason, it is still common practice to approximate $W(\omega)$ using a plasmon pole (PP) model. 
Such approach, however, may deliver an accuracy limited by its simplistic description of the frequency dependence of the polarizability, i.e. of $W$.

In this work we explore a multi-pole approach (MPA) and develop an effective representation of the frequency dependence of $W$. We show that an appropriate sampling of the polarizability in the frequency complex plane and a multi-pole interpolation can lead to a level of accuracy comparable with full-frequency methods at much lower computational cost. Moreover, both accuracy and cost are controllable by the number of poles used in MPA. Eventually we validate the MPA approach in selected prototype systems, showing that full-frequency quality results can be obtained with a limited number of poles.
\end{abstract}

\maketitle
%

\section{Introduction}
%
In the context of condensed matter physics or quantum chemistry, many body perturbation theory (MBPT) provides accurate methods to study spectroscopic properties of matter from an {\it ab initio} perspective~\cite{Onida2002RMP,martin2016book,Marzari2021NatureMat}.  
The calculations often adopt the so-called $GW$ approximation~\cite{Hedin1965PR,Aryasetiawan1998RPP,martin2016book,Reining2018wcms,Golze2019FrontChem} for the evaluation of the self-energy. 
As summarized in Sec.~\ref{section:theory}, common single-step $G_0W_0$ implementations typically make use of one-particle energies and wavefunctions from previous DFT calculations to build the non interacting one-particle Green's function $G(\omega)$ and the dynamical screened interaction potential $W(\omega)$.
Next, the self-energy is evaluated via a frequency convolution of these two quantities that give the name to the approximation.
More advanced approaches include e.g. GW self-consistency treated at different levels~\cite{vanSchilfgaarde2006PRL,Kotani2007PRB,Shishkin2007PRL,Kutepov2012PRB,Kutepov2017CPC,Grumet2018PRB},
or the adoption of vertex-corrections~\cite{Shishkin2007PRL,Chen-Pasquarello2015PRB,Ren2015PRB,Maggio2017JTCT} and cumulant expansions~\cite{Guzzo2011PRL}.
A more comprehensive discussion of these aspects can be found e.g. in Refs.~\cite{Reining2018wcms,Golze2019FrontChem}.

Since its first implementations, the $GW$ approach has been
successfully applied to a wide range of systems~\cite{Reining2018wcms,Golze2019FrontChem} for the description of
quasi-particle (QP) energies and bands as measured by ARPES experiments~\cite{Damascelli2003RMP,Hedin1998PRB,Yan2010SST,Dauth2014NJP,Gatti2020PNAS}, including spectral functions~\cite{Bechstedt1994PRB,Gatti2020PNAS}, electronic satellites~\cite{vonBarth1996PRB,Guzzo2011PRL,Caruso2018PRB}, and QP lifetimes~\cite{Marini2002PRB,Palummo2015NanLett}. 
Importantly, $GW$ quasi-particle energies are also routinely used as input for absorption spectroscopy calculations within the Bethe-Salpeter approach~\cite{Albrecht1998PRL,Onida2002RMP}.
Reflecting its wide adoption, the GW method has been implemented within multiple numerical schemes~\cite{note_manycodes}, 
ranging from localized basis sets~\cite{Blase2011PRB,Ren2015PRB,Bruneval2016CPC,Wilhelm2018JPCL}, to plane waves and pseudopotentials or PAW~\cite{Shishkin2006PRB,Marini2009CPC,Gonze2009CPC,Deslippe2012CPC,Rangel2020CPC}, to all-electron approaches using LAPW~\cite{vanSchilfgaarde2006PRL,Kotani2007PRB,Friedrich2010PRB,Jiang2016PRB}, 
also allowing for cross validation and verification~\cite{vanSetten2015JCTC,Rangel2020CPC}.

Crucial to the deployment of the method,
the frequency integration in the evaluation of the $GW$ self-energy has also been addressed in different ways. 
Common implementations of the $GW$ method make use of the so-called plasmon-pole approximation (PPA)~\cite{Hybertsen1986PRB,Zhang1989PRB,Godby1989PRL,vonderLinden1988PRB,Engel1993PRB} where, besides the different technicalities, the frequency-dependency of the polarizability is simplified through an analytical model with a single pole (for positive frequencies, plus its anti-resonant match).
The PPA method has the computational advantage of greatly simplifying the self-energy evaluation, but on the other hand its accuracy may be compromised especially for systems displaying a complex frequency structure in the screened potential.
A number of alternative methods targeting a more accurate and possibly full frequency description of the self-energy exist~\cite{Golze2019FrontChem}, including: numerical evaluation of the $GW$ frequency integrals~\cite{Lee1994PRB,Marini2002PRL,Liu2015JComputPhys,Miyake2000PRB,Shishkin2006PRB};
full frequency contour-deformation (FF-CD) methods~\cite{Godby1988PRB,book_Anisimov2000,Kotani2007PRB}, taking advantage of the analytic properties of $G$ and $W$; exact integration using the analytical structure of $W$~\cite{Bruneval2016CPC}.
While these approaches are typically accurate in terms of integration, they may turn out computationally demanding 
or somehow limited in accuracy by the analytic continuation (AC) methods~\cite{Rojas1995PRL,Riegera1999CPC,Soininen2003JPCM,vanSetten2015JCTC, Liu2016PRB,Golze2019FrontChem} required by some of them.

In this work we further explore the analytic properties of the response function 
by using a multi-pole model. The results are then used to implement a new technique, referred here as the multi-pole approximation (MPA), that allows one to obtain a simplified yet accurate description of $W$ on the frequency real axis and evaluate the $GW$ self-energy in an efficient way. The MPA technique naturally bridges from the PPA to a full-frequency treatment of the $GW$ self-energy. This new approach has been implemented numerically within the Yambo code~\cite{Marini2009CPC, Sangalli2019JPCM} and tested in different materials.

The work is organized as follows: in Sec.~\ref{section:theory} we first summarize the basic equations of the $GW$ method and present the main ideas of the MPA method, in Sec.~\ref{section:samplings} we describe an optimal frequency sampling strategy to reach a good accuracy with a reduced computational cost, and finally, in Sec.~\ref{section:performance}, we show the performance of the proposed method for three prototype bulk materials: Silicon, hBN, and rutile TiO$_2$.
In Appendix~\ref{section:interpolation} we present the mathematical details of the MPA interpolation and in Appendix~\ref{section:1p-models} we discuss in detail  different plasmon-pole models and their connection.

\section{Theory: Formulation}
\label{section:theory}

\subsection{Quasi-particle energies within GW}
%

Following Hedin's equations~\cite{Hedin1965PR,Onida2002RMP}, the $GW$ approximation for the electron-electron self-energy $\Sigma$ is obtained by neglecting vertex contributions beyond the independent-particle level (for both $\Sigma$ and the irreducible polarizability $X_0$).
This leads to an expression where $\Sigma^{GW}$ is given in terms of a frequency convolution of the Green's function $G$ with the screened Coulomb potential $W$,
\begin{equation}
    \Sigma^{GW}(\omega) = 
     \frac{i}{2 \pi}\int_{-\infty}^{+\infty} 
     d\omega' e^{-i\omega'\eta}G(\omega-\omega') W(\omega'),
    \label{eq:GW}
\end{equation}
which can be seen as the first order in a perturbation expansion involving $W$ instead of the bare interaction $v$.
By expressing the independent particle irreducible polarizability $X_0$ as
\begin{equation}
    X_0(\omega) = 
     -\frac{i}{2 \pi}\int_{-\infty}^{+\infty} 
     d\omega' G(\omega+\omega') G(\omega'),
    \label{eq:X0_convolution}
\end{equation}
the screened Coulomb interaction $W$ and the dressed polarizability $X$ can be obtained from the following Dyson equation:
\begin{eqnarray}
     \label{eq:W}
     X(\omega) &=& X_0(\omega) + X_0(\omega) v X(\omega), \\
     W(\omega) &=& \varepsilon^{-1}(\omega) v = v + v X(\omega) v. 
     \nonumber
\end{eqnarray}
In the above expressions, $v$ is the bare Coulomb potential and $\varepsilon$ is the dielectric function. Importantly, the frequency dependence of $W$ comes from its correlation term $W_c=W-v=v X v$, in turn leading to the correlation part of the self-energy, $\Sigma_c$, according to Eq.~\eqref{eq:GW}.

Usually, a non-interacting Green's function $G_0$, typically the Green's function of the Kohn-Sham (KS) system, is taken as initial guess and the $GW$ self-energy is evaluated without performing further self-consistent iterations (one-shot $G_0W_0$).
Treating the self-energy as a first order perturbation to the KS problem, 
one can then compute the quasi-particle (QP) energies, $\epsilon_m^{\text{QP}}$, either by numerically solving the exact QP equation,
\begin{equation}
   \epsilon_m^{\text{QP}} = \epsilon_m^{\text{KS}} + \langle \psi_m^{\text{KS}}|\Sigma(\epsilon_m^{\text{QP}})-v_{xc}^{\text{KS}}|\psi_m^{\text{KS}} \rangle
   \label{qp_ks}
\end{equation}
or its linearized form:
\begin{equation}
   \epsilon_m^{\text{QP}} \approx \epsilon_m^{\text{KS}} + Z_m \langle \psi^{\text{KS}}|\Sigma(\epsilon_m^{\text{KS}})-v_{xc}^{\text{KS}}|\psi_m^{\text{KS}} \rangle.
   \label{qp_ks_l}
\end{equation}
In the latter expression the renormalization factor $Z_m$ is computed from the first term of the Taylor expansion of the self energy, $\Sigma$: 
\begin{equation}
   Z_m = \left[ 1-\langle \psi_m^{\text{KS}}|\frac{\partial\Sigma(\omega)}{\partial \omega}|_{\omega=\epsilon_m^{\text{KS}}}|\psi_m^{\text{KS}} \rangle \right]^{-1} .
   \label{qp_z}
\end{equation}

In practice, in order to build the self-energy and compute quasi-particle corrections at the $G_0W_0$ level, as a first step one needs to construct the polarizability $X_0$ from the knowledge of $G_0$ according to Eq.~\eqref{eq:X0_convolution}. The former is then used for the calculation of $W$.

The Lehmann representation for the bare Green's function $G_0$, computed using the KS states is written in a compact form as:
\begin{equation}
    G_0(\omega) = \sum_{m}^{N_B} P_{m} \left[ \frac{f_m}{\omega-E_{m}-i\eta} +\frac{(1-f_m)}{\omega-E_{m}+i\eta} \right],
    \label{eq:G0}
\end{equation}
where $|\psi_m\rangle$ and $E_{m}=\epsilon_m^{\text{KS}}$ are KS eigenpairs, 
$f_m$ their occupations, $P_{m}=|\psi_m\rangle\langle \psi_m|$ their projectors, 
and the sum-over-states is usually truncated at a maximum number of bands $N_B$. Eventually, the limit $\eta \to 0^+$ is taken.
By using $G_0$ in Eq.~\eqref{eq:X0_convolution}, the irreducible polarizability can be expressed as:
\begin{equation}
    X_0(\omega) = \sum_{n}^{N_T} \frac{2 \Omega^{\text{KS}}_{n} R^{\text{KS}}_{n}}{\omega^2-(\Omega^{\text{KS}}_{n})^2},
    \label{eq:X0}
\end{equation}
where $n$ runs over single particle transitions, possibly truncated 
to $N_T$ according to the number of bands included in the calculation.
Note, however, that methods avoiding the explicit sums over empty states have been developed and made available~\cite{Umari2010PRB,Giustino2010PRB,Govoni-Galli2015JCTC}. 
In Eq.~\eqref{eq:X0}, $R^{KS}_{n}$ are the transition amplitudes computed from the Kohn-Sham states, while the poles $\Omega^{KS}_{n}$ are defined as
\begin{equation}
    \Omega^{\text{KS}}_{n} = \Delta \epsilon^{\text{KS}}_{n} - i \delta,
\end{equation}
where $\Delta \epsilon^{KS}_{n} \ge 0$ and $\delta \to 0^+$ is a damping parameter that ensures the time ordering,
similarly to $\eta$ in the case of $G_0$.

Here, for the sake of simplicity, we have kept all spatial degrees of freedom implicit and have not highlighted quantum numbers such as $\mathbf{k}$ and $\mathbf{q}$ deriving from a possible translational symmetry of the system. In this respect, and using a plane-wave basis set of $\mathbf{G}$ vectors, 
$X_0(\omega)$ and $R^{\text{KS}}_n$ in Eq.~\eqref{eq:X0} would depend on the extra indexes $\mathbf{q}\mathbf{G}\mathbf{G}'$, while the index $n$ labels transitions between states $\mathbf{k},i$ and $\mathbf{k}-\mathbf{q},j$.

\subsection{Frequency integration methods in GW}
\label{section:GW_frequency}
%

In principle, the screened interaction $W(\omega)$ needs to be computed, as the solution of Eq.~\eqref{eq:W}, for all the frequencies needed to evaluate the $GW$ self-energy according to Eq.~\eqref{eq:GW}. Nevertheless, the frequency dependence of $W$ may be quite complex, 
making the evaluation of the correlation part of the $GW$ self-energy not straightforward and computationally demanding.

Early approaches~\cite{Hybertsen1986PRB,Zhang1989PRB,Godby1989PRL,vonderLinden1988PRB,Engel1993PRB} adopted the so-called plasmon-pole model, originally proposed with explicitly real poles neglecting any spectral broadening. The simplification of the structure of $X$, and also of $W$ according to Eq.~\eqref{eq:W}, that the model provides is called the plasmon-pole approximation (PPA)~\cite{Hybertsen1986PRB,Larson2013PRB}. 
The PPA has mainly two variants, one proposed by Godby and Needs (GN)~\cite{Godby1989PRL}, and the other by Hybertsen and Louie (HL)~\cite{Hybertsen1986PRB} (though more parametrizations and refinements exist~\cite{Zhang1989PRB}, for example the von der Linden-Horsch~\cite{vonderLinden1988PRB}, or Engel-Farid~\cite{Engel1993PRB}  models). 
The analytic continuation of the polarizability $X$, within the PPA is written, for each $\mathbf{q}\mathbf{G}\mathbf{G}'$ matrix element, as
\begin{equation}
    X^{\text{PP}}(z) = \frac{2 \Omega^{\text{PP}} R^{\text{PP}}}
    {z^2-(\Omega^{\text{PP}})^2},
\end{equation}
where $\Omega^{\text{PP}}$ is defined according to $\text{Re}[\Omega^{\text{PP}}] > 0$ and $\text{Im}[\Omega^{\text{PP}}] = 0^-$.
For the sake of the present work, the two approaches are summarized and compared in Appendix~\ref{section:1p-models}.

In a full frequency real-axis (FF-RA) approach, the polarizability is evaluated  considering a dense frequency grid on the real axis and the integral for  the self energy evaluation is then computed numerically~\cite{Lehmann1972,Lee1994PRB,Marini2002PRL}. Such approach requires the use of a finite damping that broadens the structure of the polarizability~\cite{Marini2002PRL}, and a large number of frequency sampling points is typically required to converge the integral.
Other numerical integration techniques for the evaluation of the response function or the $GW$ self-energy make use of quadrature rules~\cite{book_Anisimov2000,Liu2015JComputPhys}, spectral representations of the polarizability~\cite{Miyake2000PRB,Shishkin2006PRB}, or resort to Fourier transform to imaginary time to perform frequency convolutions~\cite{Rojas1995PRL,Kutepov2012PRB,Liu2016PRB,Wilhelm2018JPCL}.

Other procedures make use of imaginary-path axis integral methods in order to transform the integration on the real axis in the self-energy into an integral over an imaginary axis~\cite{Godby1988PRB,book_Anisimov2000,Kotani2007PRB}. A similar approach resorts to a contour deformation (CD) defined in the first and third quadrants of the complex plane~\cite{Oschlies1995PRB}, in order to obtain a convenient frequency path that avoids all the poles of $W$ and encompasses only the poles of $G$.  
The integration on the real axis is then replaced by a sum of the residues of the poles in the contour plus an integral on the imaginary axis. 
This integral can be addressed either numerically~\cite{Friedrich2019PRB,Duchemin2020JCTC} or with the help of multi-pole forms~\cite{Riegera1999CPC, Soininen2003JPCM} or Pad\'e approximants~\cite{vanSetten2015JCTC,Golze2019FrontChem}. Taking advantage of the time-reversal symmetry of $W$, it is possible to reduce the frequency range in which $W$ is evaluated for the self-energy integration, either on the real or the imaginary axis~\cite{Godby1988PRB,book_Anisimov2000,Kotani2007PRB}.

Other dedicated approaches are also available.
A many-pole model for the self-energy has been developed for the calculation of inelastic losses in X-ray spectroscopy~\cite{Kas2007PRB, Kas2009JPCS}. 
Full-frequency GW has also been reformulated as a frequency-independent eigenvalue problem~\cite{Bintrim2021JCP}. 
Similarly, a spectral representation of propagators in the form of a generalized sum-over-poles combined with an algorithmic inversion technique has been recently developed and applied to the homogeneous electron gas~\cite{Chiarotti2021TOBE}.
The FF-CD has been recently used jointly with analytic-continuation techniques in an all electrons scheme that adopts a sampling along both, the imaginary axis and parallel to the real axis~\cite{Duchemin2020JCTC}. 

\subsection{The multi-pole scheme}
\label{section:mpa}
%
In this work we develop a multi-pole approach to represent $W$ and evaluate the $GW$ self-energy.
Our multi-pole scheme is based on the Lehmann representation of $X$~\cite{Fetter-Walecka1971book,martin2016book,Soininen2005PS,Ismail-Beigi2010PRB}, in which  the  polarizability is written as a sum of poles.
It is important to emphasise that, contrarily to standard PPA implementations, we consider complex poles and that
the computed poles do not correspond to single particle transitions (poles of $X_0$), but are rather intended to represent plasmon excitations. Each plasmonic pole describes the envelope of a set of transitions, with a finite imaginary part corresponding to the width of the excitation.

To represent and exploit the analytic properties of the polarizability $X$, we define a complex frequency 
$
    z \equiv  \omega + i \varpi,
$
and write $X(z)$ as the sum of a finite (and small) number $n_p$ of poles:
\begin{equation}
    X^{\text{MP}}(z) = \sum_{n}^{n_p} \frac{2 \Omega_n R_n}{z^2-\Omega_n^2}.
    \label{eq:Xmp}
\end{equation}
As for PPA, $X^{\text{MP}}, R_n, \Omega_n$ also depend on the spatial indexes $\mathbf{q}\mathbf{G}\mathbf{G}'$.
Then, we determine the parameters $\Omega_n$ and $R_n$ by interpolating the polarizability $X(z)$ computed numerically on a number of frequencies that is twice the number of poles (in order to match the unknowns). This leads to a non-linear system 
of $2 n_p$ equations and variables: 
\begin{equation}
     \sum_{n=1}^{n_p} \frac{2 \Omega_n R_n}{z_j^2-\Omega_n^2} = X(z_j), \qquad j = 1, ..., 2n_p  .
    \label{eq:nonlinear}
\end{equation}
The expression in Eq.~\eqref{eq:Xmp} is at the core of the MPA approach.
In fact, once the solution of the system is known, we obtain an analytical representation of $X(z)$ over the whole complex plane, suitable to evaluate $\Sigma^{GW}$.
Indeed, by exploiting the Lehmann representation of the Green's function, Eq.~\eqref{eq:G0}, and making use of Eq.~\eqref{eq:Xmp} to evaluate $W_c$, it is possible to compute the correlation part of the $GW$ self-energy as
\begin{multline}
       \Sigma_c(\omega) = \sum_{m}^{N_B} \sum_{n}^{n_p} P_m v R_n \Bigg[ \frac{f_m}{\omega-E_{m}+\Omega_n -i\eta} + \\
        +\frac{(1-f_m)}{\omega-E_{m}-\Omega_n +i\eta} \Bigg].
    \label{eq:Sc}
\end{multline}
This expression generalizes the PPA solution to the case of a multi-pole expansion for $X(z)$, and bridges to an exact full-frequency approach when the number of poles in $X$ is increased to convergence.

Concerning the solution of the non-linear system in Eq.~\eqref{eq:nonlinear}, several approaches are possible.
While the system can be solved analytically for a small number of poles,
in general the exact solution can be accessed numerically either by mapping the non-linear problem into an equivalent system that is linear with respect to the parameters $R_n$ and $\Omega_n$, or through the Pad\'e/Thiele procedure~\cite{Pade1977JLTPhys,Lee1996PRB,Jin1999PRB}.
A detailed description of our implementations of these approaches can be found in Appendix~\ref{section:interpolation}.

For the one pole case, the analytical solution of the interpolation with 2 complex frequencies is easily obtained:
\begin{equation}
 \left\{
    \begin{aligned}
     \Omega^2 &= \frac{X(z_1)z_1^2-X(z_2)z_2^2}{X(z_1)-X(z_2)}\\
      \,\,\,\, 2 \Omega R &= - (z_1^2-z_2^2) \frac{X(z_1)X(z_2)}{X(z_1)-X(z_2)}. 
       \end{aligned}
       \right.
     \label{eq:mpa1}
\end{equation}
We have also derived the analytical solution for the case of 2 and 3 poles, which are significantly more complex and not reported here, but are nevertheless encoded in the Yambo solver.

\section{Theory: Sampling strategies
\label{section:samplings}}

An interpolation in the form of Eq.~\eqref{eq:Xmp} is independent of the chosen sampling frequencies as far as they are all different, and the number of poles in the model, $n_p$,
equals the total number of poles of the target polarizability, $N_T$. Nevertheless, in the present approach we intent to describe $X(z)$ using a number of poles much smaller than $N_T$, and therefore the representation is not unique.
We then need to understand the possible choices for the points to be used in the interpolation of $X$. In the following, we discuss how a shift of the frequency sampling from the real axis affects the structure of the polarizability together with alternative sampling strategies. Eventually we show that the sampling plays a fundamental role in achieving a good approximation of $X(z)$ with a reduced number of poles.

\subsection{The polarizability in the complex plane}
\label{section:convolution}
%
According to the Lehmann representation, the poles of the time-ordered dressed polarizability $X$, are distributed above/below the real axis (at an infinitesimal distance), in an energy range determined by the corresponding transitions. $X$ presents a very complex structure along the real axis, 
while at increasing distance from the real axis the analytic continuation of $X$ becomes smoother.
As discussed in Sec.~\ref{section:GW_frequency}, existing approaches typically sample $X(z)$ at frequencies either along the real or the imaginary axis. As an alternative we have studied samplings with components on both axes. 

Let's consider at first a sampling of $X$ along a line parallel to the real axis, but at a distance $\varpi$.
From the point of view of the distance from the poles, this sampling effectively balances the contribution of different poles, in particular those located at large (real) frequencies. Moreover, the constant shift from the real axis smooths out the frequency dependence of the polarizability and can be understood as a filter effect resulting from the convolution between the imaginary part of the polarizability computed on the real axis, $-\text{Im}[X(\omega)]{\text{sgn}(\omega)}$, and the function  $1/\pi(\omega+ i\varpi)$ with a pole in the complex frequency plane, which is a kind of Hilbert transform:
\begin{equation}
     X(z)= -\frac{1}{\pi} \int_{-\infty}^{+\infty} 
     \frac{\text{Im}[X(\omega')]{\text{sgn}(\omega')}}{\omega-\omega'+ i \varpi} d\omega'.
     \label{eq:Hilbert_trans}
\end{equation}
The larger the value of the shift, $\varpi$, the smoother the function in the convolution and the sampled polarizability $X(z)$, and therefore the fewer the poles required to model it, as also depicted in Fig.~\ref{fig:DPcsamp}.
It is interesting to note that the FF-RA method described in Sec.~\ref{section:GW_frequency} makes typically use of a finite damping to obtain a similar simplification of the structure of $X$. In this respect, the multi-pole interpolation method presented here has the advantage that, once the parameters $R_n$ and $\Omega_n$ are obtained, it is then possible to perform a sort of deconvolution towards the real-axis, by evaluating there the polarizability when performing the integral of the self-energy. This allows us to get rid of (or at least reduce) the effect of the artificial smoothing of $X(z)$, which is not possible in the FF-RA scheme.

\subsection{Analysis of one-pole solutions}
%
Before presenting our numerical results for different sampling strategies, we discuss some analytical results useful to guide our analysis.

As shown in Appendix~\ref{section:pp_equivalence}, the two most used versions of the PPA can be mapped into a $X$ interpolated on two different frequency samplings. In fact, in the Godby-Needs (GN) scheme the parameters of the plasmon pole model (PPM) are obtained by computing $X(z)$ at $z_1=0$ and $z_2=i \varpi_p$, being $ \varpi_p$ comparable with the experimental plasma frequency of the material, while the conditions imposed by the Hybertsen-Louie (HL) scheme are shown in App.~\ref{section:pp_equivalence} to be equivalent to sampling $X$ at $z_1=0$ and $z_2= \infty$ (meaning that $X$ and $X^{\text{HL}}$ have the same leading order coefficient in the $1/z^2$ term for $z\to\infty$). 
In practice, the two methods give slightly different results: the HL-PPM tends to overestimate the position of the pole with respect to GN-PPM~\cite{Stankovski2011PRB, Larson2013PRB}, and GN-PPM reproduces better the polarizability with respect to full frequency calculations~\cite{Larson2013PRB}. Mathematically, this can be understood by considering that the interpolation of a function $X(z)$ with a structure is more effective when the sampling is done in a region of meaningful variation of the function, as done in the GN-PPM case.
It is therefore reasonable to expect that the choice of the sampled frequency points affects the interpolated $X$.

In the \suppinfo{} 
we perform a one-pole fitting, using the solution given by Eq.~\eqref{eq:mpa1}, to a test function (model polarizability) with two poles
\begin{equation}
    X^M(z)= 
     \frac{2\Omega_1 R_1}{z^2-\Omega_1^2}+\frac{2\Omega_2 R_2}{z^2-\Omega_2^2},
    \label{eq:testF}
\end{equation}
and compute a series expansions of the fitting parameters, $\Omega$ and $R$, for a perturbation on the reference sampling. This allows us to investigate the dependency of the MPA fit parameters on the sampling.
We conclude that it is possible to write equations for $\Omega$ and $R$ capturing the behaviour of both the GN- and HL-PPM schemes at the same time, and going from one to the other with a continuous function. The same analysis also shows that a sampling close and parallel to the real axis (see Fig.~\ref{fig:DPcsamp}, orange line except for the first point) introduces an error in $\Omega$ and $R$ that is proportional to the distance from the real axis, 
when comparing to the solution obtained when performing a sampling on the real axis. This means that, with $X$ sampled parallel and close to the real axis, it is convenient to stay as close to the real axis as possible. As shown in Sec.~\ref{section:dp_sampling}, this is not the case when using a double parallel sampling. 

In Sec. I.C of the \suppinfo{}, we define and report the $f$-factors in the expansions of $\Omega$ and $R$, and show that the test function behaves as a one-pole polarizability when $R_2/R_1 \to 0$ (one pole dominates) or $Q_2 \to Q_1$ (the two poles tend to coalesce). In these cases the solutions do not depend on the sampling, which supports the idea of obtaining a simplified description of the polarizability with a reduced number of poles in cases where some of the poles are close to each other or some of the residues are much larger than others. The sensitivity of the MPA method to the sampling will depend on the ratio of the residues and on the distance between the poles of $X$.
While we have not analytically investigated more complicated test polarizability functions or fitting models (including a larger number of poles), we have performed numerical analyses (e.g. a 3 pole function fitted on a 2 pole model) which tend to confirm the findings discussed above.

\subsection{Double parallel sampling}
\label{section:dp_sampling}
%

\begin{figure}
    \centering
    \includegraphics[width=0.40\textwidth]{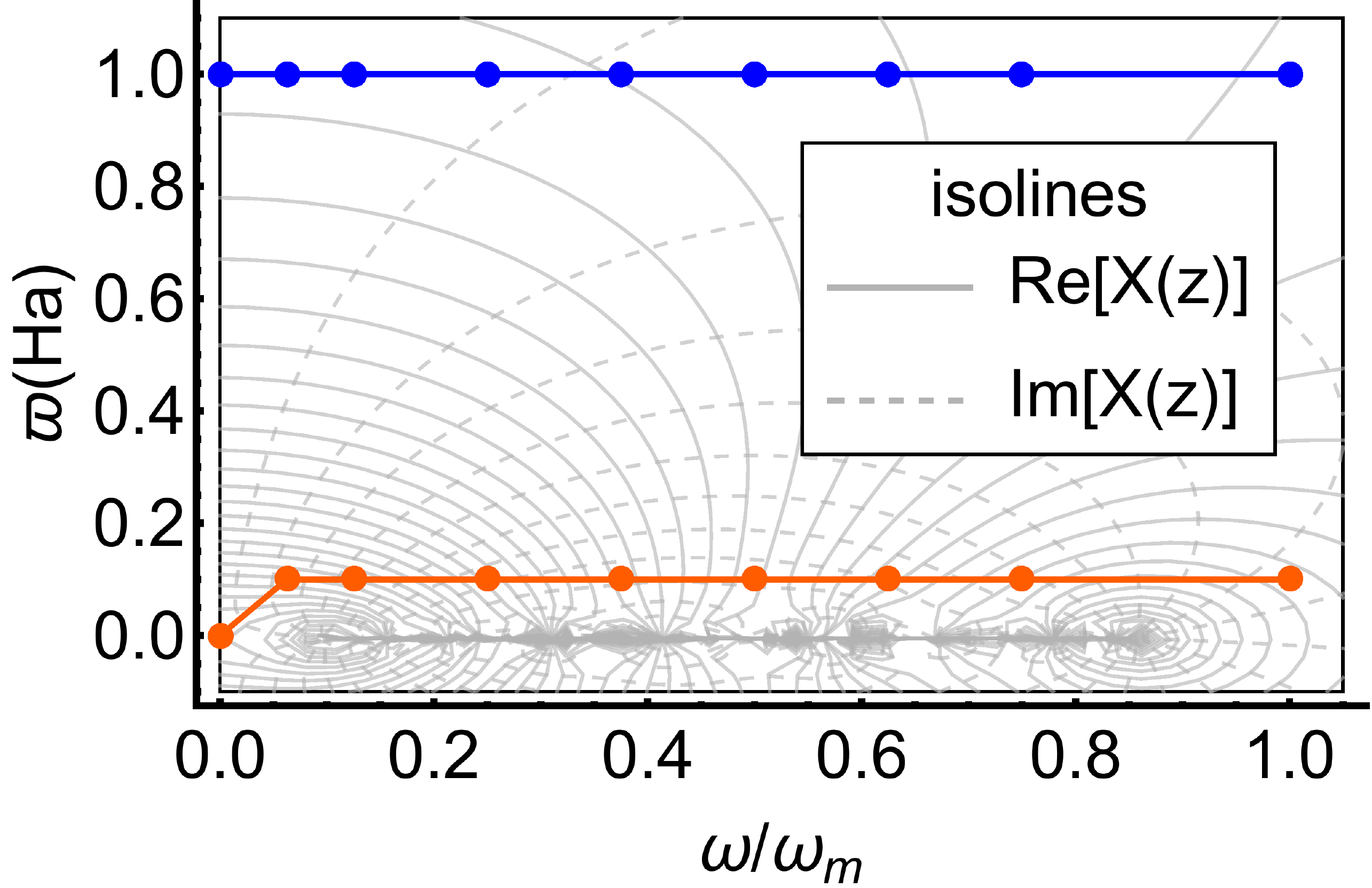}
    \caption{An illustration of the double parallel sampling with a 9 points semi-homogeneous grid along the real axis with $\varpi_2=1~Ha$, similar to the the imaginary frequency used in the GN PPM and $\varpi_1=0.1$~$Ha$, except for the origin of coordinates. The isolines in the background correspond to a toy polarizability function with 200 poles on the real axis.
    }
    \label{fig:DPcsamp}
\end{figure}

After testing several sampling strategies (samplings parallel to the real axis, tilted with a positive or negative angle with respect to it, etc), our results led to the choice of a sampling along two lines parallel to the real axis, that we will call double parallel sampling:
\begin{equation}
   s^{\text{DP}}= \left\{
    \begin{aligned}
    {\bf z^1} \text{: } z^1_n &= \omega_{n} + i \varpi_1  \\
    {\bf z^2} \text{: } z^2_n &= \omega_{n} + i \varpi_2,
    \end{aligned}
    \right.  \qquad n=1,..,n_p
    \label{eq:s_DP}
\end{equation}
where one of the two branches is a line close to the real axis while the other is located further away, e.g.  $\varpi_1<\varpi_2$. The sampling is illustrated in Fig.~\ref{fig:DPcsamp}, by the orange and green lines, while in gray we represent the isolines of a toy polarizability function with poles close to the real axis. From the isolines it is possible to see that, at some distance of the real axis, individual poles that are close enough are no longer distinguishable and contribute to a collective excitation. In a simplified view, $X$ sampled along the first line, in orange, preserves some of the structure of $X$ in the region of the poles and $X$ sampled along the second line, in blue, is simple enough to be described with a small number of poles, and accounts for the overall structure of $X$. The two branches should not be too close in order to avoid numerical instabilities arising from the  under-determination of the resulting system of equations. 

This sampling has proved to converge faster with respect to the number of poles and to be less sensitive to the distance of the first line from the real axis ($\varpi_1$) than the others we tried. This can be understood from
the analysis of a small perturbation to $z_1=0$ applied to the GN-PPM sampling (at fixed $z_2$), i.e. to the simplest one-pole double parallel sampling. 
The linear term in the perturbation of $\Omega$ and $R$ (see Eqs.~S5 and S6 in \suppinfo) 
cancels and the first perturbative term is quadratic on $\varpi$, at variance with perturbations on $z_2$ where the linear term is present (Eq.~S8 in the \suppinfo).
Moreover, considering different sampling strategies, $\Omega$ and $R$ change in the same way, when passing from a real axis sampling to a parallel or tilted sampling ($z_1=0$ only in the latter),
further stressing that overall behaviour is governed by $z_2$ and that a perturbation to $z_1$ has negligible impact.

We still have to chose a distribution of the frequencies along the real axis, $\{ \omega_n \}$, that again favours the convergence with respect to the number of poles.
Differently from the homogeneous grid used in Ref.~\cite{Duchemin2020JCTC}, we propose a partition that simply adds new frequency points when increasing the number of poles, reducing in this way oscillations in the results. 
Here we write a semi-homogeneous partition in powers of 2:
\begin{equation}
   \{\omega_{n}\}: \left\{
    \begin{aligned}
        \left(0\right) \text{, } n_p = 1  \\
        \left(0,1\right) \times \omega_m \text{, } n_p = 2  \\
        \left(0,\frac{1}{2},1\right) \times \omega_m \text{, } n_p = 3  \\
        \left( 0,\frac{1}{4},\frac{1}{2},1\right) \times \omega_m \text{, } n_p = 4  \\
        \left( 0,\frac{1}{8},\frac{1}{4},\frac{1}{2},1\right) \times \omega_m \text{, } n_p = 5   \\
        \left( 0,\frac{1}{8},\frac{1}{4},\frac{1}{2},\frac{3}{4},1\right) \times \omega_m \text{, } n_p = 6   \\
        \left( 0,\frac{1}{8},\frac{1}{4},\frac{3}{8},\frac{1}{2},\frac{3}{4},1\right) \times \omega_m \text{, } n_p = 7   \\
        ...
    \end{aligned}
    \right.
    \label{eq:w_grid}
\end{equation}
where $\omega_m$ is the extreme of the interval. We choose to use a finite value of $\omega_m$ since $X$ tends to zero for large enough frequency values, and is enough to describe its tail. 
Also supporting this option is the fact that the fulfillment of the $f$-sum rule, the condition used in the HL-PPM that is equivalent to taking $z_2= \infty$ in the GN-PPM recipe (Appendix~\ref{section:pp_equivalence}), is not critical for obtaining an accurate description of polarizability matrix elements within FF methods~\cite{Miglio2012EPJB}. The maximum value of $\omega_m$ corresponds to the largest energy transition according to the number of empty states included in the calculation of $X$. Alternatively, one can simply use a frequency with a sufficiently large real part so that it is located in the tail of the polarizability and use the other sampling points to describe its structure closer to the imaginary axis.

In practice, we use the value $\varpi_2$ used in the GN approach described in Section~\ref{section:1p-models}, in order to have the same sampling on the imaginary axis when using only one pole and a straightforward extension along the real axis when using more poles. Regarding $\varpi_1$, in case of one pole we take a null value consistently with PP models, 
while a (small but) finite value is considered for additional sampling points along the real axis in order to avoid numerical noise.
Based on the experience and results obtained using the FF real axis method, and the analytical results discussed above (Eq.~S6 in the \suppinfo)
we increase $\varpi_1$ up to 0.1~Ha.
This is also similar the values proposed for molecules in Ref.~\cite{Duchemin2020JCTC}. 

\subsection{Failure condition}
%
Sometimes, when considering only a small number of poles, the interpolation gives rise to poles that are either not physical or not reasonable, posing representability problems. This is usually solved by reassigning the values of the poles. 
An example is given by the treatment of the so-called ``unfulfilled modes" that plays an important role in different PPM schemes~\cite{Rangel2020CPC}. Here we discuss the case of
the GN-PPM approach as implemented in Yambo~\cite{Marini2009CPC,Rangel2020CPC,Godby1989PRL,Oschlies1995PRB}. 
The condition used to identify unfulfilled modes in the GN-PPM is the following:
\begin{equation}
    \text{Re}\left[ \frac{ X_{\mathbf{G}\mathbf{G}'}(\mathbf{q},0)}{X_{\mathbf{G}\mathbf{G}'}(\mathbf{q},i \varpi_p)} -1 \right] <0 \,\, \to \,\, \Omega^{\text{GN}} = 1 \text{ Ha},
    \label{eq:PPcond}
\end{equation}
where the position of the pole is set to $\Omega^{GN}=1$~Ha in case of failure. This condition is related to the estimate of $\Omega^{\text{GN}}$ according to Eq.~\eqref{eq:GNppa}.
For diagonal elements ($\mathbf{G}=\mathbf{G}'$), the polarizability evaluated on the imaginary axis is real, making the term in the square-root of Eq.~\eqref{eq:GNppa} also real.
Unfulfilled modes 
are then those for which the radicand is negative and the resulting pole imaginary.
The same condition is also used for off-diagonal matrix elements in order to ensure that the pole $\Omega^{\text{GN}}$, which is anyway taken real, 
mainly derives from the real part of the radicand in Eq.~\eqref{eq:GNppa}.

Setting the pole for unfulfilled modes at $\Omega^{\text{GN}} = 1$~Ha usually works well for semiconductors~\cite{Rangel2020CPC}, even if it may have a (usually small) impact on the quasi-particle corrections. However, in more complex systems the reliability of the PPA may be compromised either by the large number of matrix elements for which the pole is corrected or simply by the inadequacy of such a simple model correction.

In the case of MPA we propose a slightly different strategy. The condition in Eq.~\eqref{eq:PPcond}, used in the PP approach, applies to a sampling on the imaginary axis but can be generalized as:
\begin{equation}
 \Omega_n = \left\{
    \begin{aligned}
     &\sqrt{\Omega_n^2}, \qquad \qquad \text{Re}\left[ \Omega_n^2 \right] \geq 0 \\
     &\sqrt{-(\Omega_n^2)^*}, \qquad \text{Re}\left[ \Omega_n^2 \right] < 0 
    \end{aligned}
 \right.
 \label{eq:MPcond1}
\end{equation}
avoiding in this way, in case of failure, the use of a replacement constant value.
The second line in Eq.~\eqref{eq:MPcond1}, when applied, is equivalent to exchange the real and imaginary parts of $\Omega_n$. In addition, since we consider complex poles, if needed we also impose time ordering, i.e. while $\text{Re}[\Omega_n]\geq 0$ because of Eq.~\eqref{eq:MPcond1} we may force $\text{Im}[\Omega_n]<0$. 
Note this procedure is applied to all $n_p$ poles in MPA.

We now analyze the evaluation of the residues when at least one of the poles of the multi-pole interpolation is modified by the failure condition above. We start by considering a model with a single pole, for which
the residue $R$ can  be calculated using the information of either the sampling point $z_1$, first equation in Eq.~\eqref{eq:GNppa} within the GN model, or $z_2$ as
\begin{equation}
  R = \frac{X_2(z_2^2-\Omega^2)}{2\Omega} .
\end{equation}
When the pole is not corrected, the computed residue is independent of the choice between $z_1$ and $z_2$. However that is not the case if the failure condition is used. In order to improve the representation with respect to considering only one of the given solutions depending on $z_1$ or $z_2$, we propose to use Eq.~\eqref{eq:Rfit} to fit $R$.
When using more than one pole, in addition to Eq.~\eqref{eq:MPcond1} we have added an extra condition: in case a pole is close to another or its position is out of the sampling range, its residue is replaced by zero and the fit of Eq.~\eqref{eq:Rfit} is applied only to the remaining residues.

\subsection{Representability measures}
%
In order to quantify the representability error of the model with respect to the sampled points when correcting the position of the poles with Eq.~\eqref{eq:PPcond} and \eqref{eq:MPcond1}, we compute the mean number of corrected matrix elements, $\langle N_F \rangle$, and an average relative standard deviation, $\langle RSD \rangle$. 
The analysis presented in the following can be done for each $\mathbf{q}$-point, if translational symmetry is present.
For one pole, the average number of failures is simply:
\begin{equation}
    \langle N_F \rangle _{n_p=1} =\frac{1}{N_g} \sum^{N_g}_{g= {\bf GG'}}  \Theta( \Omega^{\text{MP}}_{n=1,g} ), 
\end{equation}
where $\Theta$ is a Heaviside-like step function that verify the condition:
\begin{equation}
 \Theta (\Omega) = \left\{
    \begin{aligned}
     0, \qquad \text{Re}\left[ \Omega^2 \right] \geq 0\text{ } \\
     1, \qquad \text{Re}\left[ \Omega^2 \right] < 0. 
    \end{aligned}
 \right.
 \label{eq:step}
\end{equation}
Regarding the error measurement, we use a modified version of the relative standard deviation, also known as coefficient of variation, that is then averaged over all the matrix elements. 
Since we model all the matrix elements with the same number of poles, the average deviation gathers, in a single estimate, the representability error.  
The estimator was modified by replacing in the normalization factor the mean value of the sampled values of $X$ by their maximum value. 
Namely, for 
each matrix element $g= {\bf GG'}$ (within a given $\mathbf{q}$ block, not labelled explicitly here), we define 
\begin{equation}
  X_{m g}=\max\limits_j |X_{\mathbf{G}\mathbf{G}'}(z_j)|.
\end{equation}
This is justified by the fact that $X$ is close to zero for a large region of frequencies, together with its average when it is computed with several points, which may results in an inadequate use of the coefficient of variation. 
The error estimators then read:
\begin{equation}
    \langle RSD \rangle _{n_p=1} =\frac{1}{N_g} \sum^{N_g}_{g= {\bf GG'}} \frac{1}{X_{m g}} \sqrt{ \sum^{2}_{j=1} |X^{\text{MP}}_{j g}-X_{j g}|^2 }
\end{equation}
\begin{equation}
    \langle RSD \rangle_{\text{PPA}} =\frac{1}{N_g} \sum^{N_g}_{g= {\bf GG'}} \frac{\Theta( \Omega^{\text{PP}}_{n g})}{X_{m g}} \sqrt{ \sum^{2}_{j=1} |X^{\text{PP}}_{j g}-X_{j g}|^2 } .
\end{equation}
In case of the PPA we compute the error just for the matrix elements that fail the condition in order to favour the comparison with MPA, since there may be a deviation already due to the fact of discarding the imaginary part of the poles in the PPA (MPA is based on an interpolation, while PPA is not).

In the general multi-pole case we define:
\begin{equation}
    \langle N_F \rangle =\frac{1}{N_g} \sum^{N_g}_{g={\bf GG'}} \frac{\sum^{n_p}_n \Theta (\Omega^{\text{MP}}_{n g}) |R^{\text{MP}}_{n g}|}{\sum^{n_p}_n |R^{\text{MP}}_{n g}|} 
\end{equation}
\begin{equation}
    \langle RSD \rangle =\frac{1}{N_g} \sum^{N_g}_{g= {\bf GG'}} \frac{1}{X_{m g}} \sqrt{ \frac{1}{2 n_p -1} \sum^{2 n_p}_{j=1} |X^{\text{MP}}_{j g}-X_{j g}|^2 },
    \label{eq:rsd}
\end{equation}
where we have introduced a normalization by the residues of each pole in the counter of the failure condition $\langle N_F \rangle$, in order to differentiate the contribution of each pole.   
%

\section{Results: MPA performance}
\label{section:performance}
\begin{figure*}
    \centering
    \includegraphics[width=0.95\textwidth]{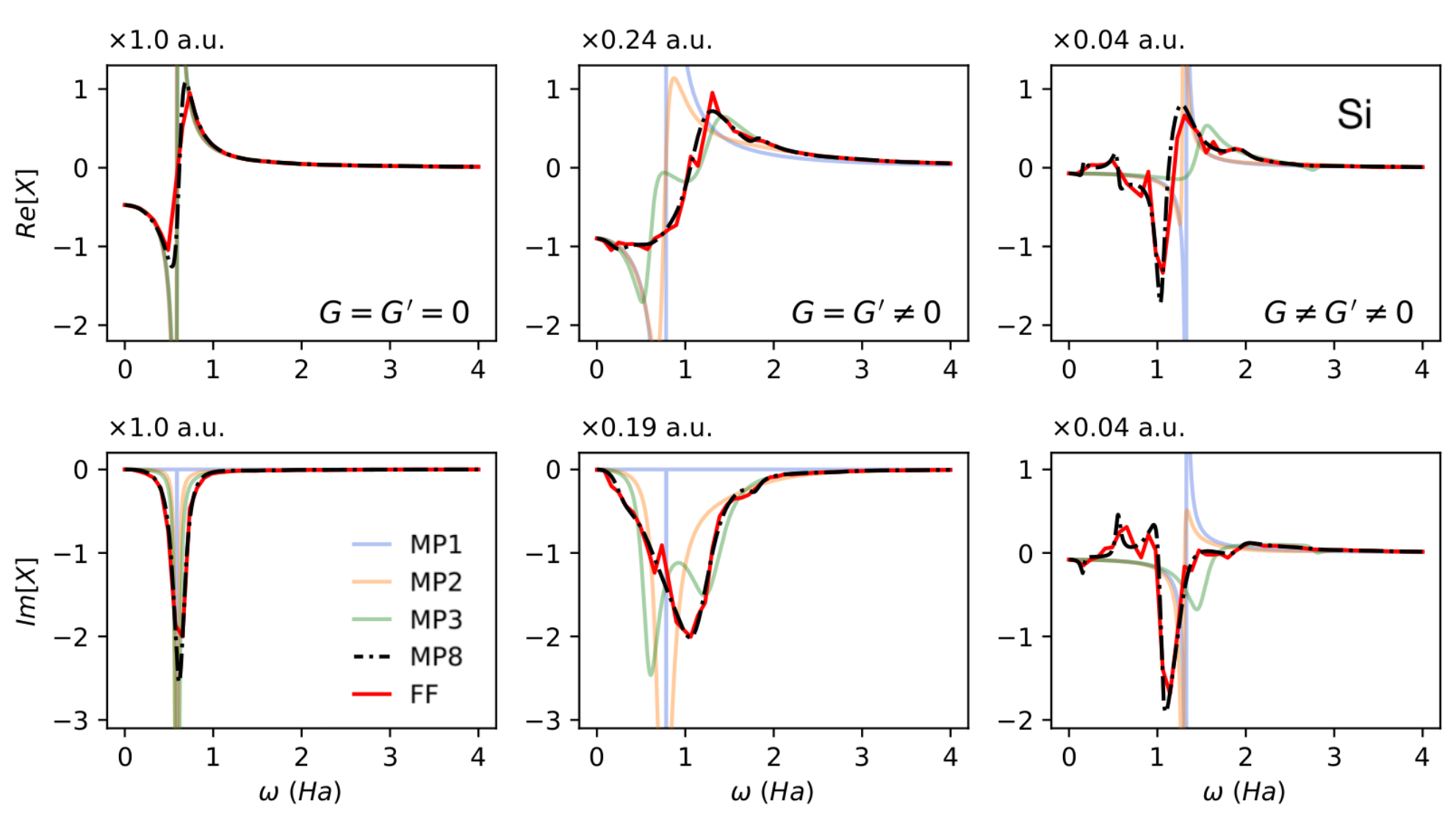}
    \caption{Selected Si $X$ matrix elements computed within MPA with 1, 2, 3 and 8 poles and compared with the corresponding FF real-axis results. Although the real and imaginary parts of the function are plotted using different arbitrary units, the different matrix elements can be compared since their scale is consistent and indicated in each plot.}
    \label{fig:Xpol_Si}
\end{figure*}
\begin{figure*}
    \centering
    \includegraphics[width=0.95\textwidth]{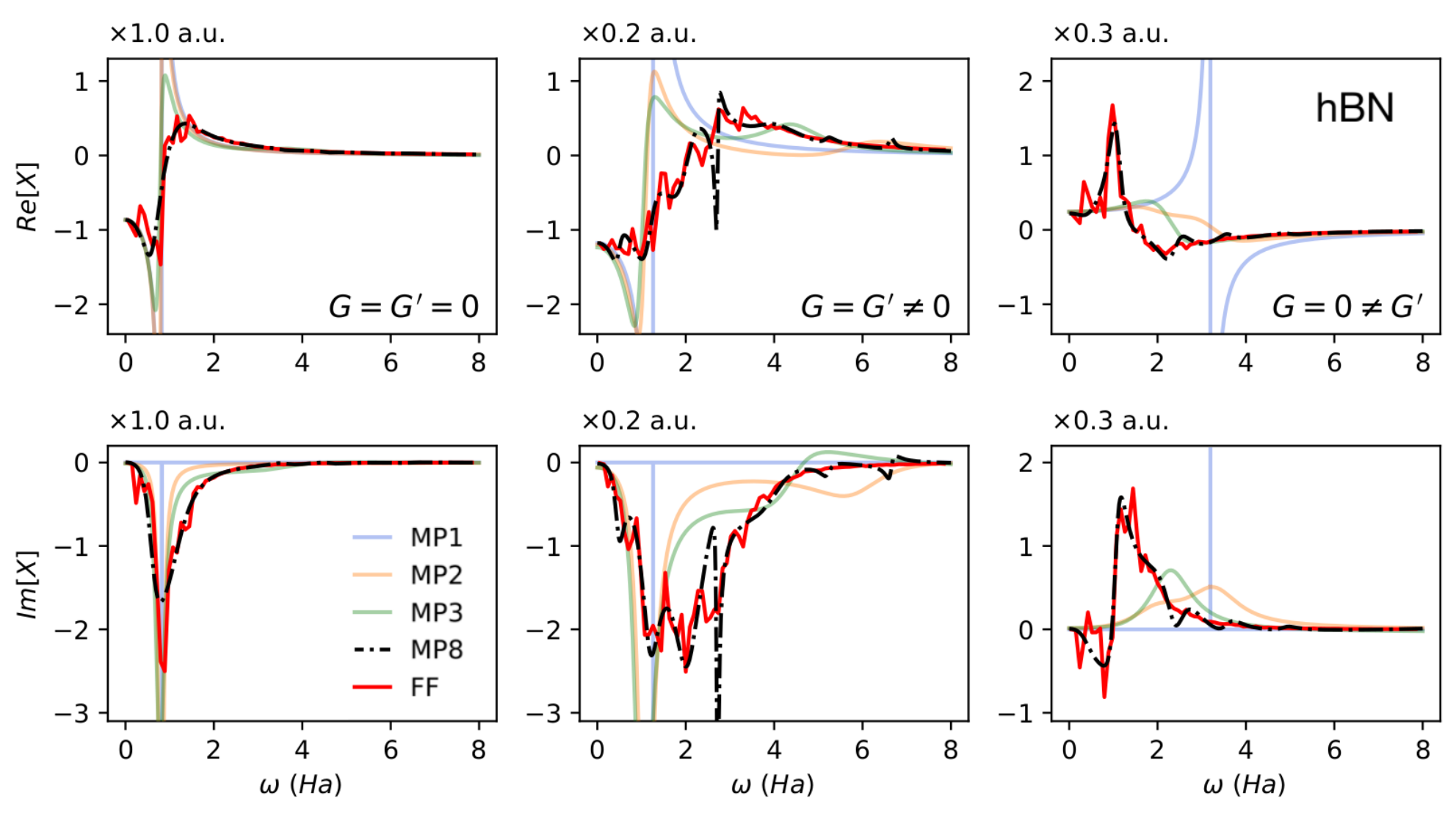}

    \caption{Selected hBN $X$ matrix elements computed within MPA with 1, 2, 3 and 8 poles and compared with the corresponding FF real-axis results. The same scheme from Fig.~\ref{fig:Xpol_Si} is used for the units.}
    \label{fig:Xpol_hBN}
\end{figure*}

\begin{figure*}
    \centering
    \includegraphics[width=0.95\textwidth]{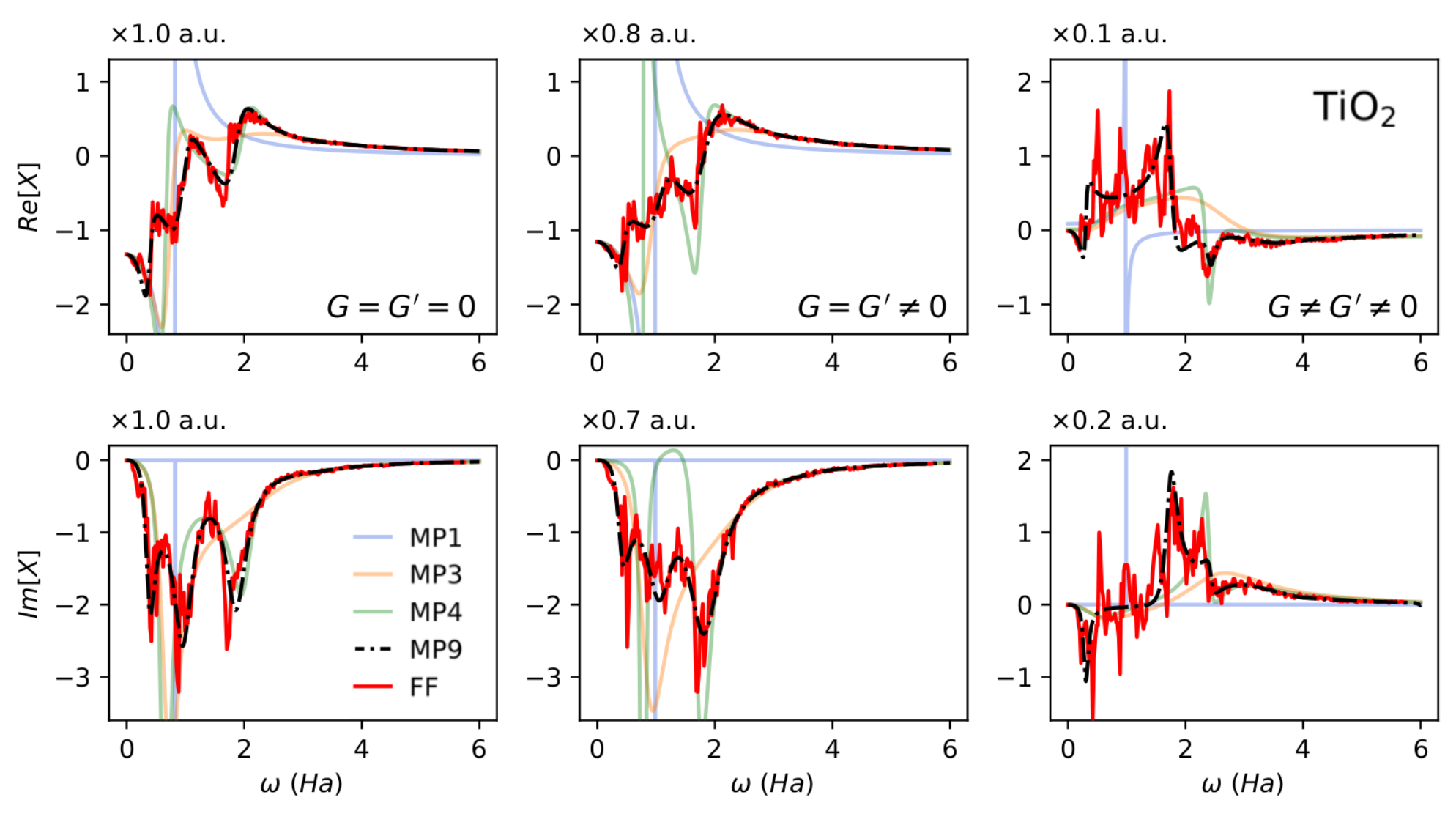}
    \caption{Selected TiO$_2$ $X$ matrix elements computed within MPA with 1, 3, 4 and 9 poles and compared with the corresponding FF real-axis results. The same scheme from Fig.~\ref{fig:Xpol_Si} is used for the units.}
    \label{fig:Xpol_TiO2}
\end{figure*}

We have validated the MPA method in three different bulk materials: Si, a prototype semiconductor, hBN with AA and AA' staking, and rutile TiO$_2$, a mid band-gap semiconductor oxide.
We compare the MPA approach described above with PPA and FF calculations and with the existing literature. In particular, we compare our results for Si with Refs.~\cite{Kutepov2009PRB, Rangel2020CPC}, hBN with Refs.~\cite{Wickramaratne2018JPCC,Mengle2019APLMater}, and  TiO$_2$ with Ref.~\cite{Rangel2020CPC}.

DFT calculations were performed using the Quantum ESPRESSO package~\cite{Giannozzi2009JPCM,Giannozzi2017JPCM}. We employed the LDA exchange-correlation functional for Si and hBN, while GGA-PBE for TiO$_2$, with norm-conserving 
pseudopotentials in all cases. 
For Si, we use a grid of $12\times12\times12$ $\mathbf{k}$-points, a kinetic energy cutoff of 20 Ry and 300 KS states to perform sums-over-states.
In case of hBN, we used $\mathbf{k}$-points meshes of $18\times18\times9$ and $18\times18\times6$, corresponding to AA and AA' stacking respectively, with an energy cutoff of 60 Ry and 400 KS states.
For rutile TiO$_2$, we use a shifted $\mathbf{k}$-grid of $4\times 4 \times 6$ $\mathbf{k}$-points, a kinetic energy cutoff of 70 Ry for the wavefunctions,
and 600 KS states. 

GW calculations were performed with the Yambo code~\cite{Marini2009CPC,Sangalli2019JPCM}. We use a standard Monte Carlo stochastic scheme called Random Integration Method (RIM)~\cite{Marini2009CPC,Rozzi2006PRB} to treat integrals over the Brillouin zone with Coulomb divergence.  
The RIM technique is used in order to accelerate convergences with respect to the $\mathbf{k}$-point mesh in case of hBN and TiO$_2$, but not for Si simply to illustrate that the MPA works well independently of this choice. The size of the polarizability matrix is set to 25, 10 and 15~Ry for Si, hBN and TiO$_2$, respectively. 
In the case of hBN, the value of 10~Ry is not sufficient to converge  quasi-particle corrections, and a more suitable value is 25~Ry. Due to the high computational cost of FF real-axis calculations, we decided to perform the comparison of PPA, MPA, and FF methods for hBN with a X matrix cutoff of 10~Ry, 
while the results obtained with 25~Ry (quasi-particle energies in the case of PPA and MPA) are given in Table~\ref{tab:QPs}.
Regarding the self-energy evaluation, in the bare Green function we use a damping parameter $\eta=0.1$~eV for all calculations.

\subsection{The polarizability matrix}
%

\begin{figure}
    \centering
    \includegraphics[width=0.4\textwidth]{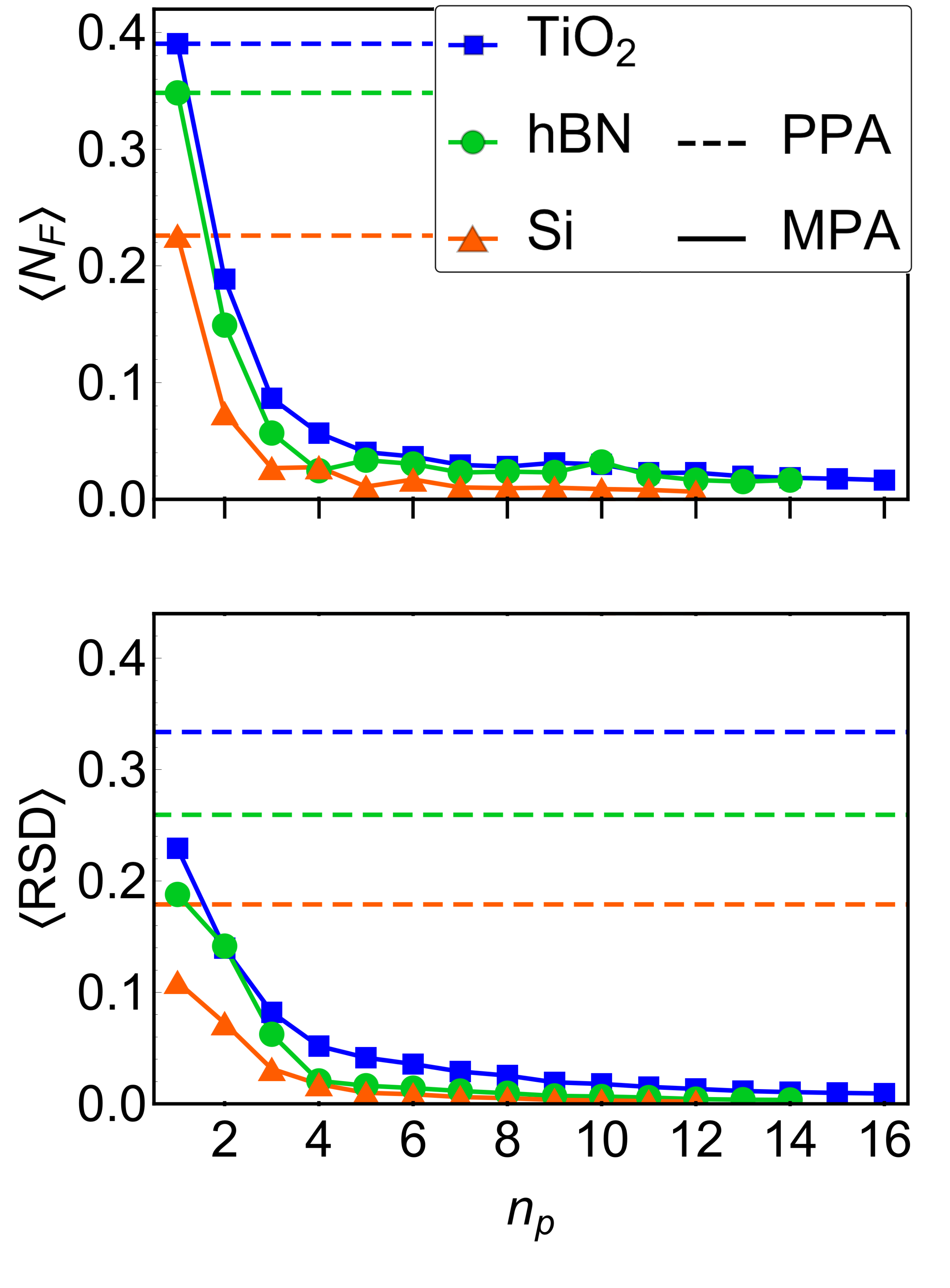}
    \caption{(Top) $\langle N_F \rangle$, mean number of matrix elements for which the position of the poles was corrected according to Eq.~\eqref{eq:MPcond1} for MPA and Eq.~\eqref{eq:PPcond} for PPA, and (bottom) $\langle RSD \rangle$, average deviation as defined in Eq.~\eqref{eq:rsd}, as a function of the number of poles used in the MPA approach. Due to the high computational cost of the FF calculations, in case of hBN the comparison is made with a non fully converged dimension of the polarizability matrix, of 10~Ry. The plots correspond to the AA stacking, but similar values are seen also for the AA' stacking with 10~Ry of X matrix.}
    \label{fig:repro_t}
\end{figure}

In Figs.~\ref{fig:Xpol_Si}, \ref{fig:Xpol_hBN} and \ref{fig:Xpol_TiO2} we plot  a set of diagonal and off-diagonal matrix elements of the polarizability for three different bulk materials: Si, hBN, and TiO$_2$ rutile, respectively. 
Each plot shows the real and imaginary parts of the polarizability computed within MPA using a different number of poles, compared with the corresponding FF real-axis results. The same number of poles 
is used for all the matrix elements. 
In fact, on the one hand the multiple peak structures of $X$ is more complex 
for large ${\bf G}$-vectors, 
while on the other hand their maximum amplitude, and therefore their weight in the integration of the self-energy, decreases with ${\bf G}$. This means that a more simplistic description of the structure of these elements will not affect much the computed $\Sigma$.
For reasons of computational convenience, besides the number of poles, the array of sampled frequencies is also the same for all the matrix elements.
Of course, the complexity of $X$ depends on the material under study. In silicon, for instance, the most important matrix elements have an almost single-peak structure that favours the use of a single pole, while for TiO$_2$ the first element, ${\bf G}= {\bf G'}=0$, already shows several peaks and a slower decay of the maximum amplitude with respect to ${\bf G}$. 

The imaginary part of diagonal elements of $X$ describes the spectral properties of the polarizability, and therefore is always negative with peaks around the real part of the poles and widths given by their imaginary part. A model with a single pole describes approximately the envelop of the real part of diagonal elements, but is unable to describe the width of the main peak, since the small value of the imaginary part of the pole obtained from the interpolation is translated into a delta-like peak. The use of a complex pole is not that different from a real one obtained by neglecting its imaginary part, as in the PPA case. The inclusion of a second pole however, may improve the description of the imaginary part of the diagonal elements of $X$ even if there are no significant improvements in the real part.
On the other hand, the off-diagonal matrix elements of $X$ show a mixture of their real and imaginary parts, since the residues of the poles in Eq.~\eqref{eq:X0} have an imaginary part depending on their ${\bf G}$ and ${\bf G'}$ components. Off-diagonal matrix elements are more likely to fail the PP condition of Eq.~\eqref{eq:PPcond} and even a single but complex pole and the generalized failure condition of Eq.~\eqref{eq:MPcond1} may lead to considerable improvements on the description of $X$, as shown in the right panels of Fig.~\ref{fig:Xpol_Si}. 

In general the description of the polarizability improves quickly as the number of poles increases, as demonstrated by the representability measurements $\langle N_F \rangle$ and $\langle RSD \rangle$ plotted in Fig.~\ref{fig:repro_t}. Already with one pole, the error of the generalized condition is lower than the one obtained with the PPM and then both $\langle N_F \rangle$ and $\langle RSD \rangle$ rapidly decrease with increasing number of poles.

\subsection{The GW quasi-particle correction}
%

\begin{figure}
    \centering
    \includegraphics[width=0.48\textwidth]{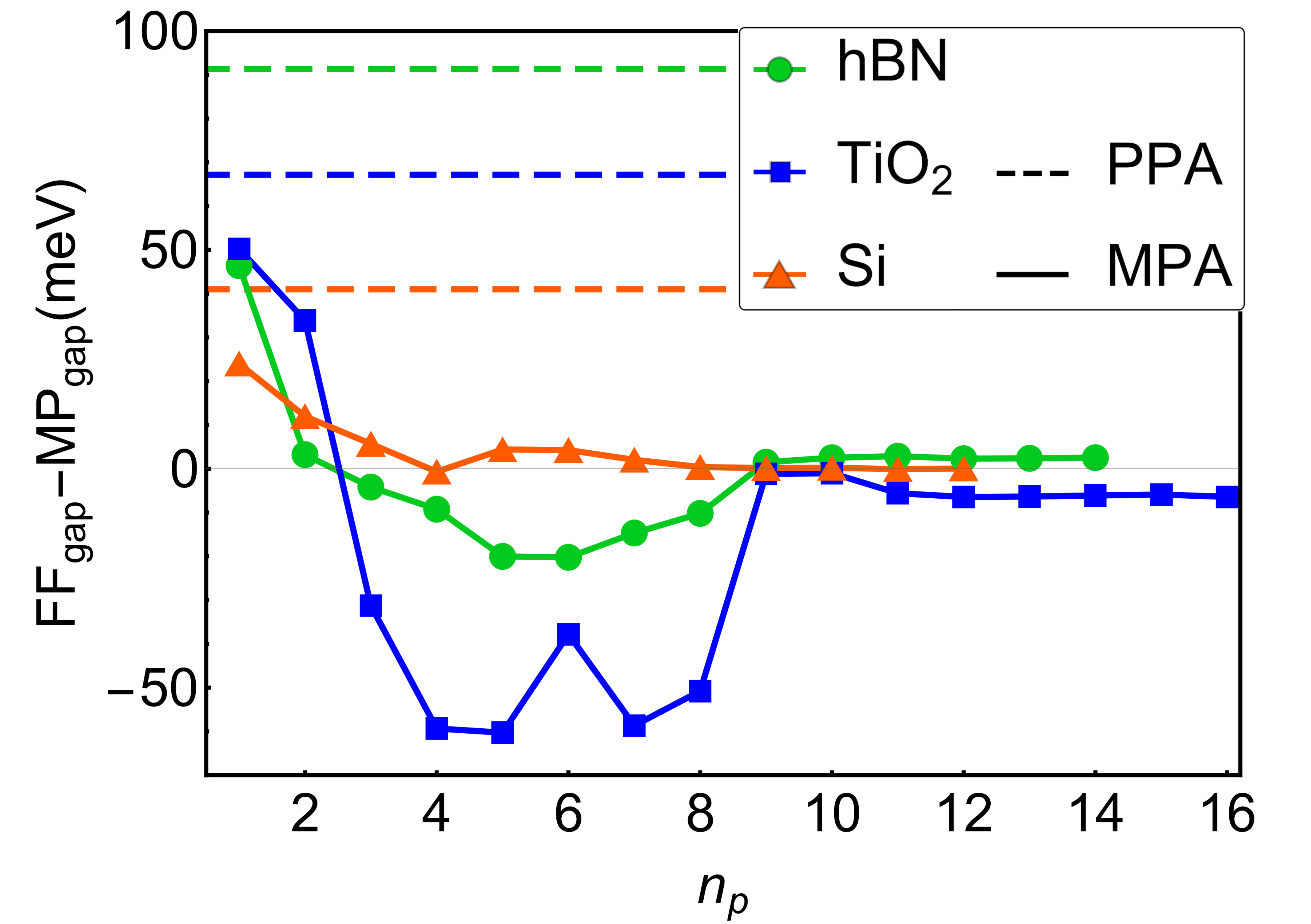}
    \caption{Deviations of the fundamental gap calculated via the MPA with respect to FF results as a function of number of poles for Si, hBN and TiO$_2$. As in Fig.~\ref{fig:repro_t}, in case of hBN we use 10~Ry of X matrix  and plot only the results of the AA stacking.  }
    \label{fig:delta_gaps_t}
\end{figure}

In Fig.~\ref{fig:delta_gaps_t} we report the convergence of the GW correction to the band gap with respect to the number of poles used in the MPA for the three systems under study. 
The same frequency sampling was used in the three cases, namely a double parallel sampling, Eq.~\eqref{eq:s_DP} with the grid given by Eq.~\eqref{eq:w_grid} along the real axis and shifts $\varpi_1=0.1$ Ha and $\varpi_2=$~1 Ha.
For TiO$_2$, which has a more structured polarizability, the quasi-particle corrections are more sensitive to the sampling. With a single pole model using different values of $\varpi_2$ the results can differ by as much as $\sim 100$~meV. In the multi-pole case the difference decreases to $6$~meV 
when comparing results obtained with $\varpi_2=0.5$ and 1.0~Ha. Even if we expect the highest value to lead to a slightly simpler $X$, the convergence is reached with the same number of poles in both cases.

Interestingly, the same convergence behaviour is found for the three systems: between 8 and 11 poles the quasi-particle corrections differ by less than $1$~meV from the FF results. The number of poles needed to obtain convergence is much more homogeneous than the number of frequencies in FF real axis calculations (300 for Si, 400 for hBN and 1500 for TiO2), since the shifts $\varpi$ in the double parallel sampling determine the structure of the polarizability and therefore the number of poles required to model it.  
\begin{table}
  \begin{ruledtabular}
  \begin{tabular}{l|cccccc}
  %
 {\bf System}&   {\bf QP(eV)}                 && {\bf PPA}  & {\bf MPA}   && {\bf FF}   \\[4pt]
  \hline\\[-3pt] 
   {\bf Si} &            $\Gamma_c \to \Gamma_v$     && 3.28    & 3.30   && 3.30  \\ 
     &            $K_c      \to \Gamma_v$     && 1.26    & 1.30   && 1.30  \\[5pt] 
   \hline\\[-3pt]
   {\bf hBN$_{AA}$} &    $M_c      \to K_v$          && 7.35    & 7.47   && 7.46 [7.25]  \\ 
     &            $K_c      \to K_v$          && 7.02    & 7.16   && 7.16 [6.91]  \\ 
     &            $H_c      \to K_v$          && 5.33    & 5.50   && 5.50 [5.23]  \\
     &            $L_c      \to K_v$          && 5.26    & 5.42   && 5.42 [5.17]  \\[5pt] 
   \hline\\[-3pt]
   {\bf hBN$_{AA'}$} &   $L_c      \to K_v$          && 6.21    & 6.24   && 6.24 [6.20]  \\ 
     &            $K_c      \to K_v$          && 6.17    & 6.20   && 6.20 [6.17]  \\ 
     &            $H_c      \to K_v$          && 6.02    & 6.05   && 6.05 [6.01]  \\
     &            $M_c      \to K_v$          && 5.93    & 5.98   && 5.98 [5.92]  \\[5pt] 
   \hline\\[-3pt]
   {\bf TiO$_{2}$} &     $\Gamma_c \to \Gamma_v$     && 3.20    & 3.27   && 3.26  \\[5pt]
 \end{tabular}
 \end{ruledtabular}
 \caption{Quasi-particle (QP) transitions computed with the linearized QP Eq.~\eqref{qp_ks_l} on top of PP, MPA and FF. In case of hBN, the numbers in brackets correspond to the actual FF calculations done with 10~Ry of X matrix, these values are shifted outside the brackets according to the differences observed between the MPA results with 10 and 25~Ry respectively.} \label{tab:QPs}
\end{table} 

We report our final QP results in Table~\ref{tab:QPs}. In the case of hBN, the difference between the PPA and FF-RA values changes with respect to the size of the polarizability matrix. At 10~Ry we found similar differences for stacking AA and AA' with a maximum value of 0.10~eV. At 25~Ry the difference increases up to a maximum value of 0.17~eV for the stacking AA and decreases up to a maximum value of 0.05~eV for the stacking AA', the plasmon-pole approximation being more accurate in the AA' configuration.

\subsection{Self Energy and Spectral Function}
%
\begin{figure}
    \centering
    \includegraphics[width=0.48\textwidth]{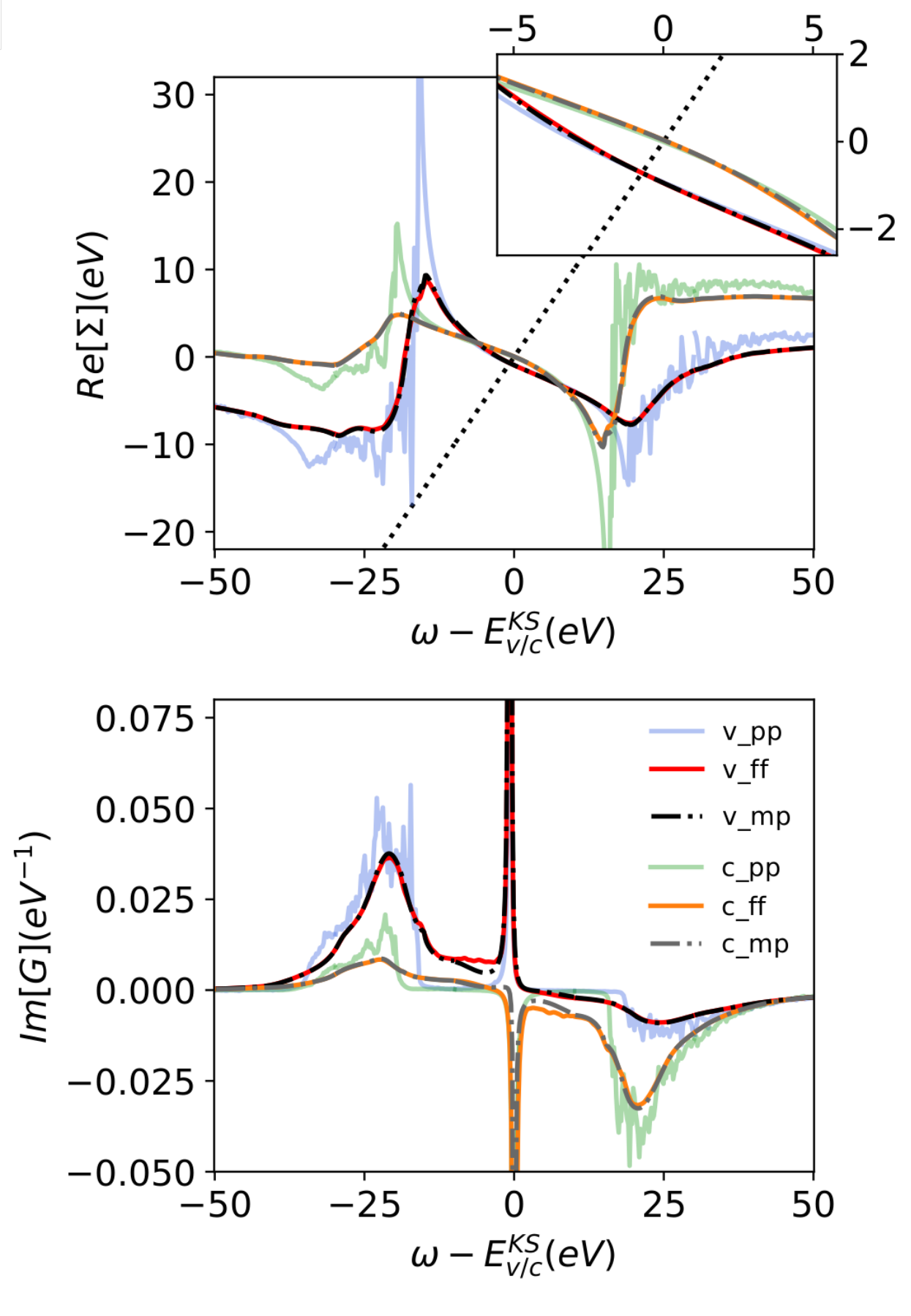}
    \caption{Frequency dependence of the Si valence (v) and conduction (c) real part of the self-energy (top) and spectral function (bottom) of the quasiparticles involved in the fundamental gap computed with PPA, MPA and FF.}
    \label{fig:Sg_Si}
\end{figure}
The present MPA method targets the dynamical dependence of the dressed polarizability, in turn directly related to $W$, in order to perform the frequency integral in the self-energy and finally to compute the quasiparticle corrections to the independent particle states. For this purpose, we need a good description of $\Sigma(\omega)$ in a frequency region around the solution of the quasiparticle equation, Eq.~\eqref{qp_ks}. However, the description of the self energy in a larger range of frequencies is interesting by itself, since $\Sigma(\omega)$ contains, besides the quasiparticle energies, information about many-body features like satellites and quasiparticle lifetimes. These properties are computed by solving the corresponding Dyson equation for the Green function, $G=G_0+G_0\Sigma G$. Through Eq.~\eqref{eq:Sc}, the MPA method provides a description for the frequency dependence of the self-energy, that we assess in this Section. 

In Fig.~\ref{fig:Sg_Si} we compare the Si self energy (top) and spectral function (bottom), obtained as the imaginary part of the dressed Green's function, $\text{Im}\left[ G \right]$, computed within PPA, MPA, and FF-RA. 
The self-energy presents a typical two-pole structure~\cite{Rojas1995PRL, Riegera1999CPC} corresponding to the contributions of the empty and occupied states.
The picture can be better understood taking Eq.~\eqref{eq:Sc} into account: occupied states contribute to the self-energy at energies around the value $-\Omega$ plus their own KS energies (negative), different from the empty states that contribute at $+\Omega$ plus a positive term from the KS energies further separating the two contributions of the main plasmon at $\pm \Omega$.

The PPA gives a good description of the tail of the main peaks, especially including the region around the solution of the quasiparticle equation, and the overall behaviour of $\Sigma$. However, intrinsic representability problems appear due to the inability of PPA in describing the imaginary part of $W$. The result is a very noisy $\Sigma$, particularly around its peaks, giving rise to noisy satellite peaks in the spectral function. 
In contrast, the MPA self energy (computed with 8 poles, i.e. the number required to obtained converged quasiparticle energies) reproduces quite well the FF-RA results in the whole energy range.
In fact, there is a reversal on the level of complexity of the polarizability with respect to the self-energy: $X$ is smooth when computed within PPA, and very structured in FF-RA and MPA, whereas $\Sigma$ is very spiky within PPA and smooth for FF-RA and MPA.  

This result can be understood by analysing the effect of 
the Dyson equation for $W$, Eq.~\eqref{eq:W},
on the polarizability matrix elements $X_{\bf GG'}({\bf q}, \omega)$. At the independent particle level, $X_0$ presents a large number of 
peaks described by Eq.~\eqref{eq:X0}.
When applying the Dyson equation, the poles from different matrix elements of $X_0$ are combined, resulting in a dressed polarizability $X$ with broader peaks corresponding to plasmonic excitations, as shown in Figs~\ref{fig:Xpol_Si}, \ref{fig:Xpol_hBN} and \ref{fig:Xpol_TiO2}.
Thanks to this feature, it is possible to use few (complex) poles in the MPA modeling (then approximated with 1 real pole in PPA) of each matrix element $X_{\bf GG'}({\bf q},\omega)$. 
However, there is a further pole superposition of the matrix elements in the integral of the self-energy (Eq.~\eqref{eq:Sc}), whose accurate description in a full frequency range needs a proper modeling of the imaginary part of $X$. This superposition in PPA is partially remedied by the finite $\eta$ damping of the Green function, while the presence of finite imaginary parts in the MPA poles naturally improves the PPA description.
In Silicon, the proximity of the several electronic states and the number of additional low intensity plasmon-excitations result in a noisy behaviour of the PPA self-energy around the main peaks. In contrast, the FF and MPA methods properly describe the superposition of all the excitations, and the resulting $\Sigma(\omega)$ function is smooth.

This picture may be different for less screened systems like e.g. molecules~\cite{Lischner2014PRB,vanSetten2015JCTC}, where the energy levels are scattered, the plasmonic 
excitations are far apart, and the plasmon-pole model may lead to a self-energy that presents a simpler structure than a FF treatment. On the other hand, the nonexistence of a gap in metallic systems may lead to a self-energy with a single main peak~\cite{Cazzaniga2012PRB} also noisy within PPA. However, the special cases of metals with small plasmon energies or strongly correlated materials are particularly challenging for simple models like the PPA, since the quasi-particle solutions lie in a zone of multiple plasmonic excitations, as shown in Ref.~\cite{Miyake2013PRB} for SrVO$_3$. The applicability of the PPA on such systems has been discussed in the literature~\cite{Aryasetiawan1998RPP,Marini2002PRL} and the advantages in using the MPA method for a metallic case are the subjects of a future work.

\section{Conclusions}
\label{section:conclusion}
%
In this paper we have introduced and developed a multi-pole approximation (MPA) to represent the reducible polarizability and to evaluate the correlation self-energy in the GW method. The MPA method can be seen as a generalization of the commonly used plasmon-pole approximation (PPA), while increasing the number of poles in the description of $X$ it tends to the exact, full frequency solution. Therefore, MPA naturally bridges from PPA to FF GW, with controllable accuracy and computational cost.
We have provided numerical methods to compute the MPA parameters and investigated in detail the effect of  different frequency samplings on the procedure. In doing so, we have also discussed two common formulations of the PPA (Godby-Needs and Hybertsen-Louie PPM's) showing how they can be seen in a unified frame.
Eventually, the MPA method has been validated and benchmarked on selected bulk semiconductors (Si, hBN, rutile TiO$_2$), showing systematic improvement over the PPA, and numerical agreement with FF GW already with about 10 poles.

The present MPA approach, in the few poles regime, considerably improves the quasi-particle energies with respect to PPA without a significant increase in the computational cost. When considering more poles, around 10 for the systems studied here, its computational cost is comparable to AC approaches. In this regime we have shown that MPA reaches an accuracy comparable with standard FF contour deformation methods~\cite{vanSetten2015JCTC, Golze2019FrontChem}, significantly more demanding. Moreover, we believe MPA presents several advantages with respect to this method.
In the contour deformation approach, the number of frequencies used in the evaluation of $W$ increases with the distance of the state from the Fermi level due to the increasing number of poles of $G$ entering in the contour~\cite{Duchemin2020JCTC}, whereas within MPA all the quasi-particles have the same computational cost. 
Another advantage of MPA relies on its analytic form, which allows one to solve analytically the frequency integral of the self-energy that has also a multi-pole form. Moreover, the frequency structure of the polarizability is meaningful and permits the analysis of the plasmonic interactions. The MPA technique applies irrespectively of the basis set, and can also be straightforwardly extended beyond $G_0W_0$, e.g. to  quasiparticle self-consistent $GW$ approaches~\cite{vanSchilfgaarde2006PRL,Kotani2007PRB}.

Overall, our findings show that the multi-pole approach can be used to obtain a simple and effective representation of response functions.
We illustrated how to analyse simple types of sampling in order to understand and design good recipes.
We show that the MPA with optimal sampling strategies in the complex plane can lead to a level of accuracy comparable with full-frequency methods at much lower costs,
not only for the quasi-particle energies, but also for the whole energy range relevant for the self-energy. 

\section*{Acknowledgments}
%
We acknowledge stimulating discussions with Pino D'Amico, Alberto Guandalini, Miki Bonacci, Simone Vacondio and Matteo Zanfrognini.  This work was partially supported by the MaX -- MAterials design at the eXascale -- European Centre of Excellence, funded by the European Union program H2020-INFRAEDI-2018-1 (Grant No. 824143).
We also thank SUPER (Supercomputing Unified Platform -- Emilia-Romagna) from Emilia-Romagna POR-FESR 2014-2020 regional funds.
Computational time on the Marconi100 machine at CINECA was provided by the Italian ISCRA program.

\begin{appendix}
%
\section{MPA interpolation} 
\label{section:interpolation}

\subsection{Linear solver
\label{section:linear_solver}}

The non-linearity of the system of equations in Eq.~\eqref{eq:nonlinear} depends only on inverse factors involving the variables $\Omega_n$, since the system is otherwise linear in $R_n$. It is possible to separate these two behaviours by following the procedure described below. We start by splitting the sampled points $\{z_j, X(z_j)\}$ in two sets, for example:
\begin{equation}
\begin{aligned}
    s_1: \text{ } & j = 1, ..., n_p, \\
    s_2: \text{ } & j = n_p+1, ..., 2n_p .
\end{aligned}
 \label{eq:split}
\end{equation}
The first set defines a matrix $\mathbf{A_1}$ and vector $\mathbf{x_1}$ as follows:
\begin{eqnarray}
     \mathbf{A_1}_{mn} &=& \frac{2 \Omega_n}{z_m^2-\Omega_n^2}, \\
     \mathbf{x_1}_{m} &=& X(z_m), \qquad
    n, m = 1, ..., n_p
    \nonumber
\end{eqnarray}
such that we can write a linear system for the vector $\mathbf{r}= ( R_1, R_2, ..., R_{n_p})$:
\begin{equation}
    \mathbf{A_1} \mathbf{r} = \mathbf{x_1} .
    \label{eq:sys1}
\end{equation}
We can do the same with the other half of the data, by defining the matrix $\mathbf{A_2}$ and the vector $\mathbf{x_2}$ as:
\begin{eqnarray}
     \mathbf{A_2}_{mn} &=& \frac{2 \Omega_n}{z_{n_p+m}^2-\Omega_n^2}, \\
     \mathbf{x_2}_{m} &=& X(z_{n_p+m}), 
    \qquad n, m = n_p+1, ..., 2n_p
    \nonumber
\end{eqnarray}
leading to
\begin{equation}
    \mathbf{A_2} \mathbf{r} = \mathbf{x_2} .
    \label{eq:sys2}
\end{equation}
Either Eq.~\eqref{eq:sys1} or \eqref{eq:sys2} can be used to compute the residues if the positions of the poles are known. Furthermore, from these two equations it is possible to obtain a complete set of $n_p$ equations for $\Omega_n$:
\begin{equation}
    \mathbf{r}=(\mathbf{A_1})^{-1}\mathbf{x_1}=(\mathbf{A_2})^{-1}\mathbf{x_2} .
\end{equation}
Here we explore in depth this idea using a different formulation that maps the problem into an equivalent linear system that can be easily solved with standard linear algebra tools. 

Within the multipole model, $X(z)$ can be written in the form of a particular Pad\'e approximant, i.e. as a fraction of two polynomials $N(z^2)$ and $D(z^2)$ of degree $n_p-1$ and $n_p$, respectively:
\begin{multline}
   X(z)=\frac{N_{n_p-1}(z^2)}{D_{n_p}(z^2)}=\\
       =\frac{a_1+a_2z^2+...+a_{n_p}z^{2(n_p-1)}}{b_1+b_2z^2+...+b_{n_p}z^{2(n_p-1)}+z^{2n_p}},
    \label{eq:pade}
\end{multline}
where the factorization of $D_{np}(z^2)$ gives  the position of all the poles:
\begin{equation}
   D_{n_p}(z^2)=\prod_{n=1}^{n_p} (z^2-\Omega_n^2) .
    \label{eq:D}
\end{equation}

On the other hand, the coefficients $a_n$ in the numerator involve combinations of poles and residues. From the point of view of the system of equations, we are changing the unknown variables $\{R_n,\Omega_n\}$ into $\{a_n,b_n\}$. If we consider a vector constructed from the progression of powers of $z$: ${\bf Z}(z)=(1,z^2, ..., z^{2(n_p-1)})$, and vectors ${\bf a}$ and ${\bf b}$ constructed from the coefficients of $N(z^2)$ and $D(z^2)$: ${\bf a}=(a_1,a_2, ..., a_{n_p})$ and ${\bf b}=(b_1,b_2, ..., b_{n_p})$, we can write Eq.~\eqref{eq:pade} in a more compact way and then make it linear with respect to the new variables:
\begin{eqnarray}
   & &X(z)=\frac{{\bf Z}(z)\cdot{\bf a}}{{\bf Z}(z)\cdot{\bf b}+z^{2n_p}}
    \label{X}
   \\[5pt]
   & &X(z)z^{2n_p} + X(z)\, {\bf Z}(z)\cdot{\bf b}=
   {\bf Z}(z)\cdot{\bf a}  .
    \label{eq:ab_z}
\end{eqnarray}

By using the first set of points of Eq.~\eqref{eq:split} we can define the following vector and matrices:
\begin{eqnarray}
 \mathbf{v_1}&=&
 \begin{bmatrix}
 X(z_1)z_1^{2n_p}\\
 X(z_2)z_2^{2n_p}\\
 \vdots\\
 X(z_{n_p}){z_{n_p}}^{2n_p}\\
 \end{bmatrix} 
    \label{eq:v1}
\\[8pt]
 \mathbf{Z_1}&=&
 \begin{bmatrix}
 1&z_1^2&...&z_1^{2(n_p-1)}\\
 1&z_2^2&...&z_2^{2(n_p-1)}\\
 \vdots&\vdots&& \vdots\\
 1&z_{n_p}^2&...&z_{n_p}^{2(n_p-1)}\\
 \end{bmatrix}
    \label{eq:Z1}
\\[8pt]
 \mathbf{M_1}&=&
 \begin{bmatrix}
 X(z_1)&X(z_1)z_1^2&...&X(z_1)z_1^{2(n_p-1)}\\
 X(z_2)&X(z_2)z_2^2&...&X(z_2)z_2^{2(n_p-1)}\\
 .&.&&.\\
 .&.&&.\\
 .&.&&.\\
 X(z_{n_p})& X(z_{n_p})z_{n_p}^2&...& X(z_{n_p})z_{n_p}^{2(n_p-1)}\\
 \end{bmatrix}
 \,\,\,\,\,\,\,\,\,\,\,\,
    \label{eq:M1}
\end{eqnarray}
$\mathbf{Z_1}$ is known as a square Vandermonde matrix~\cite{Ycart2012arxiv} and is always invertible as far as all the sampling points are different~\cite{Macon1958TAMM}.
Then we can write Eq.~\eqref{eq:ab_z} in matrix form:
\begin{equation}
 {\bf v_1}+{\bf M_1}{\bf b}=
{\bf Z_1}{\bf a},
    \label{eq:ab_1}
\end{equation}
while an analog equation is obtained with the second set of points: 
\begin{equation}
 {\bf v_2}+{\bf M_2}{\bf b}=
{\bf Z_2}{\bf a} .
    \label{eq:ab_2}
\end{equation}
By combining Eq.~\eqref{eq:ab_1} and ~\eqref{eq:ab_2} to remove vector ${\bf a}$ we obtain a linear system for $\bf b$: 
\begin{eqnarray}
 {\bf M}{\bf b} &=& {\bf v}
  \label{eq:b}
  \\[7pt]
  {\bf M} &=& \mathbf{Z_1}^{-1}\mathbf{M_1}-\mathbf{Z_2}^{-1}\mathbf{M_2} 
   \label{eq:Mv} \\
  {\bf v} &=&  -\mathbf{Z_1}^{-1}\mathbf{v_1}+\mathbf{Z_2}^{-1}\mathbf{v_2} .
    \nonumber 
\end{eqnarray}
In terms of the number of matrix inversions to be performed, Eqs.~\eqref{eq:b}-\eqref{eq:Mv} can be recast into the following more practical form:
\begin{equation}
 \left[ \mathbf{Z_2} \mathbf{Z_1}^{-1} \mathbf{M_1}- \mathbf{M_2} \right]{\bf b}=
 -\mathbf{Z_2} \mathbf{Z_1}^{-1} \mathbf{v_1}+\mathbf{v_2}
    \label{eq:b2}
\end{equation}

Before we proceed, in order to the reduce numerical instabilities of the algorithm, 
it is convenient to carry out a normalization 
\begin{eqnarray}
    z_j&=&y_j z_{\text{max}}, \\
    z_{\text{max}}&=&\max(|z_n|),
     \qquad n=1,..,n_p, 
    \nonumber
\end{eqnarray}
where, i.e., $z_{\text{max}}$ is
the largest frequency modulus in the sampling set involved with the matrix we need to invert, ${\bf Z_1}$.

The proposed normalization balances the large differences among the matrix elements that emerge when increasing the number of poles, due to the range of the sampling and the increasing powers. 
Rescaled unknowns can then be defined as:
\begin{equation}
 \left\{ 
      \begin{aligned}
      a'_n=a_n (z_{\text{max}})^{n-1} \\
      b'_n=b_n (z_{\text{max}})^{n-1}
      \end{aligned}
   \right.
  \label{eq:ab_p}
\end{equation}
Likewise, we also define the $\mathbf{Y_{1,2}}$ and $\mathbf{M'_{1,2}}$ matrices similarly to $\mathbf{Z_{1,2}}$ and $\mathbf{M_{1,2}}$, Eqs.~\eqref{eq:Z1} and~\eqref{eq:M1} respectively, by using $y_n$ instead of $z_n$.
Eventually, Eqs.~\eqref{eq:b}-\eqref{eq:Mv} can be re-written as a linear system for $\bf b'$:
\begin{equation}
   \left[ \mathbf{Y_2} \mathbf{Y_1}^{-1} \mathbf{M'_1}- \mathbf{M'_2} \right] \mathbf{b'} =
     - \mathbf{Y_2} \mathbf{Y_1}^{-1} \mathbf{v_1}+ \mathbf{v_2}
    \label{eq:bp2}
\end{equation}

Once we have computed $\mathbf{b}'$, and in turn $\bf b$ via Eq.~\eqref{eq:ab_p}, we need to obtain the poles $\Omega_n$, which are the variables with physical meaning and those required in the evaluation of the self-energy integral. As mentioned above, this is equivalent to find the zeros of the polynomial $D_{n_p}(z^2)$ in Eq.~\eqref{eq:D}, and a powerful method to perform this task is the diagonalization of the corresponding companion matrix:
\begin{equation}
 \mathbf{C}=
 \begin{bmatrix}
 0&0&...&0&-b_1\\
 1&0&...&0&-b_2\\
 0&1&...&0&-b_3\\
 \vdots&\vdots&\ddots&\vdots&\vdots\\
 0&0&...&1&-b_{n_p}\\
 \end{bmatrix}
    \label{eq:CM}
\end{equation}
The eigenvalues of $\bf C$ correspond to the squares of the position of the poles, $\Omega_n^2$. 

As mentioned before, in case all the poles are different, 
the residues may be computed with either Eq.~\eqref{eq:sys1} or \eqref{eq:sys2}. Alternatively we can use all the $2n_p$ points to fit $R_n$ with a linear least squares method:
\begin{equation}
   \min\limits_{\mathbf{r}} ||\mathbf{A} \mathbf{r} - \mathbf{x}||,
    \label{eq:Rfit}
\end{equation}
where, for each $n = 1, ..., n_p$, and $j = 1, ..., 2n_p$, one has
\begin{eqnarray}
     \mathbf{A}_{jn} &=& \frac{2 \Omega_n}{z_{j}^2-\Omega_n^2}, 
     \qquad
     \mathbf{x}_{j} = X(z_{j}) 
\end{eqnarray}
A similar leas-square approach was also recently adopted in the definition of the sum-over-poles approach~\cite{Chiarotti2021TOBE}, and applied to the case of the homogeneous electron gas.

\subsection{Pad\'e/Thiele solver}
\label{section:pade}
%

An alternative way to the method described above consists in solving at the same time both polynomials $N$ and $D$ in the Pad\'e interpolant in Eq.~\eqref{eq:pade} by means of the Thiele's interpolation formula, witch expresses the interpolant as a 
continued fraction of the reciprocal differences~\cite{Pade1977JLTPhys}. The number of required steps corresponds to the number of points to be interpolated, that is $2n_p$ in our case: 
\begin{equation}
   \frac{N(z^2)}{D(z^2)}=\frac{c_1}{1+}\frac{c_{2}(z^2-z_1^2)}{1+}...\frac{c_{2n_p}(z^2-z_{2n_p-1}^2)}{1+(z^2-z_{2n_p-1}^2)g_{2n_p}(z)}, 
    \label{eq:thiele}
\end{equation}
where the coefficients $c_s$ and functions $g_s(z)$ are given by the following recursion relations:
\begin{eqnarray}
  c_s&=&g_s(z_s)
  \label{eq:recursion_c}
  \\[7pt]
  g_s(z) &=& \left\{
  \begin{aligned}
  & X(z_s), \qquad \qquad s=1 \\
  & \frac{g_{s-1}(z_{s-1})-g_{s-1}(z)}{(z-z_{s-1})g_{s-1}(z)},  s\geq 2,      
  \end{aligned}
  \right.
    \label{eq:recursion_g}
\end{eqnarray}
where index $s=1,..,2n_p$ represents both the iteration step and the index of the corresponding point in the given set.
The polynomials $N(z^2)$ and $D(z^2)$ and their coefficients can then be computed recursively. Notice that in Thiele's relation the definition of each polynomial differs from the one in Eq.~(\ref{eq:pade}) by a multiplicative constant that, however, does not affect the zeros of the polynomials, and moreover cancels out in the fraction.   

As in the case of the linear solver, with Thiele's procedure we are interested in computing only the monomial coefficients of the denominator, since we can always obtain the residues by means of Eqs.~\eqref{eq:sys1} or~\eqref{eq:Rfit}.
The recipe~\cite{Pade1977JLTPhys} for the polynomial in the denominator is:
\begin{equation}
 D_s(z)= \left\{
\begin{aligned}
   & 1, \qquad \qquad \qquad s=0,1 \\
   & D_{s-1}(z)-c_s(z-z_{s-1})D_{s-2}(z), \quad s\geq 2.
\end{aligned}
  \right.
    \label{eq:recursion_D}
\end{equation}
The desired polynomial of degree $n_p$ is obtained in the last step, $s=2n_p$. Notice that in this notation the index $s$ does not reflect anymore the degree of the polynomial.
In each iteration the degree of the polynomial $D_s(z)$ is $(s-1)/2$ for odd integers and $s/2$ for even numbers. 
We can write a vector of the coefficients of the final polynomial, ${\bf d}=(d_1,...,d_{n_p},d_{n_p+1})$. It has an extra dimension comparing to vector ${\bf b}$ in Eq.~(\ref{eq:ab_z}) due to the multiplicative constant mentioned before that goes to the higher order monomial of $D$ in Eq.~(\ref{eq:pade}). 
The recursion of the polynomial in Eq.~(\ref{eq:recursion_D}) can be easily translated into a recursion of the vector ${\bf d}$:
\begin{equation}
 {\bf d^s} = \left\{
  \begin{aligned}
   & (1,0,..,0), \qquad \qquad  s=0,1 \\
   & d^s_i=d^{s-1}_i+c_s z_{s}d^{s-2}_{i+1}-c_s z_{s-1}d^{s-2}_i,  s\geq 2,
  \end{aligned}
   \right.
    \label{eq:recursion_vec_d}
\end{equation}
where the second term of the sum in the rhs of Eq.~(\ref{eq:recursion_vec_d}) is computed just for $i=1,...,n_p$, while the other two include also the last dimension, $i=n_p+1$.

The recursion of the $c_s$ coefficients can also be recast in vectorial form, by considering in each iteration, $s$, a $2n_p$ dimensional vector, ${\bf c}$:
\begin{equation}
 {\bf c^s}: \left\{
  \begin{aligned}
   (X(z_1),..,X(z_{2n_p})) \text{, } s=1\text{ } \\
    c^s_j=\frac{c^{s-1}_{j-1}-c^{s-1}_{j}}{(z_{j}-z_{j-1})c^{s-1}_{j}} \text{, } s\geq 2;
  \end{aligned}
   \right.
    \label{eq:recursion_vec_c}
\end{equation}

\noindent
where now the iterations are represented by index $s$ in the upper position, and the vector coordinates and point identifiers are represented by index $j$ in the lower position. 
Notice however that only one particular component of vector ${\bf c^s}$ enters in Eq.~\eqref{eq:recursion_vec_d} at each iteration, i.e. $c_s\equiv c^s_s$, other components
are worth only to update the vector.
Once computed ${\bf d}$, the relation with the coefficient of vector ${\bf b}$ is as simple as $b_n=d_n/d_{n+1}$, that recovers the unitary coefficient accompanying the higher order monomial of the polynomial. The position of the poles can then be computed with the companion matrix, Eq.~\eqref{eq:CM}, likewise in the case of the linear solver. 

\section{Plasmon pole models
\label{section:1p-models}}
%
As stated in Sec.~\ref{section:GW_frequency},
there are several flavours of the plasmon-pole model (PPM), that are compared for instance in Refs.~\cite{Larson2013PRB,Shaltaf2008PRL}, but two of them are more commonly used, one due to Hybertsen and Louie~\cite{Hybertsen1986PRB} (HL), and the other due to Godby and Needs~\cite{Godby1988PRB} (GN). In the Hybertsen and Louie (HL) approach, two physical constraints are imposed: (1) compliance with the Kramers-Kronig relations in the limit of small frequencies, and (2) compliance with the $f$-sum rule.
The Kramers-Kronig relations for $X^{\text{HL}}$ provide:
\begin{eqnarray}
  \label{eq:KK_HL}
  X^{\text{HL}}(0) &=& \frac{2}{\pi} \mathcal{P}\!\!\int_0^\infty {d\omega \, \omega^{-1}  \text{Im}[X^{\text{HL}}(\omega)]} \\
  &=& \frac{2}{\pi} \mathcal{P}\!\!\int_0^\infty {d\omega \, \omega^{-1}  \text{Im}[X(\omega)]} = X(0),
  \nonumber
\end{eqnarray}
whereas the $f$-sum rule is enforced by the condition:
\begin{equation}
  \frac{2}{\pi} \int_0^\infty {d\omega \, \omega  \, \text{Im}[X^{\text{HL}}(\omega)]}= 
\frac{2}{\pi} \int_0^\infty {d\omega \, \omega \,  \text{Im}[X(\omega)]} = S,
\end{equation}
where $X$ is the computed dressed polarizability.
In the above equation, the result of the integral ($S$) can be expressed in terms of some of the components of the electronic density and Coulomb interaction.
On a plane-wave basis, it reads:
%
%
\begin{eqnarray}
    \nonumber
  S_{\mathbf{ {\bf GG'}}}(\mathbf{q}) &=& -\frac{\omega_p^2}{v(\mathbf{q}+\mathbf{G})} \frac{\rho(\mathbf{G}-\mathbf{G'})}{\rho(0)}
  \frac{(\mathbf{q}+\mathbf{G})\cdot(\mathbf{q}+\mathbf{G'})}{|\mathbf{q}+\mathbf{G}|^2} \\[5pt]
   &=& \rho(\mathbf{G}-\mathbf{G'}) \, \left[ 
   (\mathbf{q}+\mathbf{G})\cdot(\mathbf{q}+\mathbf{G'})\right]
\end{eqnarray}
where $\rho$ is the electronic density and the plasma frequency, $\omega_p$, is computed as $\omega_p=\sqrt{4 \pi \rho(0)}$.

In the Godby and Needs (GN) approach, the polarizability is evaluated at two different frequencies located along the imaginary axis of the frequency plane: $z=0$ and $z=i \varpi_p$, being $ \varpi_p$ comparable with the plasma frequency of the material. 
\begin{equation}
    \label{eq:KK_GN}
   ( R^{\text{GN}}, \Omega^{\text{GN}}): \left\{
    \begin{aligned}
    & X^{\text{GN}}(0) =  X(0)  \\
    & X^{\text{GN}}(i \varpi_p) =  X(i \varpi_p).
    \end{aligned}
    \right.
\end{equation}
Other versions of the PPA are based on the above recipes 
aiming at improving the description of the off-diagonal matrix elements of the polarizability
~\cite{vonderLinden1988PRB,Engel1993PRB,Giantomassi2011PSSB}.  
We will now take a closer look at the GN and HL approaches, from the point of view of the multi-pole approximation.

\subsection{Connecting the GN- and HL-PPM schemes} \label{section:pp_equivalence}
%
The idea of interpolating the two parameters $(R^{\text{PP}},\Omega^{\text{PP}})$ of the plasmon-pole model 
starting from $X$ evaluated at two different frequencies, used by the GN-PPM, is very flexible, and multiple sampling options can be adopted.
Nevertheless, using different pairs of sampling frequencies typically leads to different parametrization of the resulting PPM. 
With this in mind, we search for the pair of frequency points to be used in Eq.~\eqref{eq:mpa1} that would correspond to the conditions imposed by the HL scheme. 
The equations for the GN-PPM are:
\begin{eqnarray}
    \label{eq:GNppa}
    & &X(0) =  - \frac{2 R^{\text{GN}}}{\Omega^{\text{GN}}}   \\
    & &
     \Omega^{\text{GN}} = \varpi_p \, \text{Re}\left[ \frac{ X(i \varpi_p)}{X(0) - X(i \varpi_p)} \right]^{\frac{1}{2}},
    \nonumber
\end{eqnarray}
while those for the HL-PPM are:
%
%
\begin{eqnarray}
     \label{eq:HLppa}
    & &X(0) =  - \frac{2 R^{\text{HL}}}{\Omega^{\text{HL}}}   \\
    & &2 \Omega^{\text{HL}} R^{\text{HL}} =  S.
    \nonumber
\end{eqnarray}
We note that Eq.~\eqref{eq:KK_HL} in the HL formulation, connected to the Kramers-Kronig relation, implies the first condition (equality of $X$ and $X^{\text{HL}}$ at $\omega=0$) imposed in the GN recipe, Eq.~\eqref{eq:KK_GN}, as also evident in comparing Eqs.~\eqref{eq:HLppa} and~\eqref{eq:GNppa}.

If we consider the exact polarizability $X$, written in the Lehmann representation similar to the multi-pole model given in Eq.~\eqref{eq:Xmp},
but with all the $N_T$ addends and with $\text{Im}[\Omega_n] \to 0^-$, and solve the integral in the $f$-sum rule we get:
%
%
\begin{equation}
     \frac{2}{\pi} \int_0^\infty {d\omega \, \omega\,  \text{Im}[X(\omega)]} = 2\sum_n {\Omega_{n}} R_{n}.
\end{equation}
Then, the condition imposed by HL in the plasmon-pole model is  
%
%
\begin{equation}
{2 \Omega^{\text{HL}}} R^{\text{HL}} = {2}\sum_n {\Omega_{n}} R_{n}.
\end{equation}
This relation 
imposes the equality of the $1/z^2$ coefficient (leading order) of the  asymptotic behaviour of the polarizabilities, making explicit the known  fact that sum rules describe properties at infinity~\cite{Crowley2015arxiv}:
\begin{equation}
\lim_{\omega \to \infty} X^{\text{HL}}(\omega) / X(\omega) = 1 .
\label{eq:HL_tail}
\end{equation}
This means that in the long frequency range, $\omega \gg \max\limits_n |\Omega_n|$, $X$ behaves as a one-pole function with the exact same asymptotic decay of $X^{\text{HL}}$. 
Thus, we can think of the HL representation as a limiting case of GN-PPM recipe when the second frequency goes to infinity:
\begin{equation}
   ( R^{\text{HL}}, \Omega^{\text{HL}}): \left\{
    \begin{aligned}
     & X^{\text{HL}}(0) =  X(0)  \\
     & X^{\text{HL}}(\infty) = X(\infty),
    \end{aligned}
    \right.
\end{equation}
where the evaluation of the infinite frequency in the second equation is taken as the limit of the $X^{\text{HL}}/X$ ratio, according to Eq.~\eqref{eq:HL_tail}.

\end{appendix}

\bibliography{paper_MPA_resub}

\end{document}


\title{Frequency dependence in GW made simple using a multi-pole approximation: Supplemental Material}
\author{Dario A. Leon$^{1,2}$, Claudia Cardoso$^{2}$, Tommaso Chiarotti$^{3}$, Daniele Varsano$^{2}$, Elisa Molinari$^{1,2}$, and Andrea Ferretti$^2$}
\affiliation{$^1$FIM Department, University of Modena \& Reggio Emilia, Via Campi 213/a, Modena (Italy)}
\affiliation{$^2$S3 Centre, Istituto Nanoscienze, CNR, Via Campi 213/a, Modena (Italy)}
\affiliation{$^3$Theory and Simulation of Materials (THEOS), Ecole Polytechnique F\'ed\'erale de Lausanne (EPFL), CH-1015 Lausanne (Switzerland)}

\maketitle
%


\section{Analysis of different samplings
\label{section:analysis}}
%
In the following, we consider the following test function (modelling a polarizability with two pairs of poles)
\begin{equation}
    y(z)= x_1(z) + x_2(z) \equiv  \frac{2\Omega_1 R_1}{z^2-\Omega_1^2}+\frac{2\Omega_2 R_2}{z^2-\Omega_2^2}
    \label{eq:testF}
\end{equation}
and study how different sampling options affect the solution of the fitting procedure when using one pole.

\subsection{Perturbing the GN-PPM sampling}
%
Let us define the reference GN sampling with an imaginary frequency of modulus $\varpi_0$:
\begin{equation}
   s^{GN}: \left\{
    \begin{aligned}
    z_1 &= 0  \\
    z_2 &= i \varpi_0
    \end{aligned}
    \right.
\end{equation}
and those we want to compare with, starting with a perturbation on $z_1$:
\begin{equation}
   s^{0}: \left\{
    \begin{aligned}
    z_1 &= \omega  \\
    z_2 &= i \varpi_0
    \end{aligned}
    \right.
\end{equation}
\begin{equation}
   s^{i0}: \left\{
    \begin{aligned}
    z_1 &= i \varpi  \\
    z_2 &= i \varpi_0,
    \end{aligned}
    \right.
\end{equation}
where $\varpi<\varpi_0$. 
When $\omega$ and  $\varpi$ are small, we have a similar behaviour with the same parameters:
\begin{equation}
\left\{
    \begin{aligned}
   \Omega^{0} &= \Omega^{GN} \left(1-\frac{f_1^{0}}{\varpi_0^2} \omega^2 + O[\omega^3] \right) \\
   R^{0} &= R^{GN} \left(1-\frac{c_1^{0} f_1^{0}}{\varpi_0^2} \omega^2 + O[\omega^3] \right) 
    \end{aligned}
    \label{eq:samp_0->0}
    \right.
\end{equation}
\begin{equation}
\left\{
    \begin{aligned}
   \Omega^{i0} &= \Omega^{GN} \left(1+\frac{f_1^{0}}{\varpi_0^2} \varpi^2 + O[\varpi^3] \right) \\
   R^{i0} &= R^{GN} \left(1+\frac{c_1^{0} f_1^{0}}{\varpi_0^2} \varpi^2 + O[\varpi^3] \right). 
    \end{aligned}
    \label{eq:samp_i0->0}
    \right.
\end{equation}
We also consider a perturbation on $z_2$ as:
\begin{equation}
   s^{\varpi_0}: \left\{
    \begin{aligned}
    z_1 = 0\text{ }  \\
    z_2 = i (\varpi_0+\varpi).
    \end{aligned}
    \right.
    \label{eq:s_w0}
\end{equation}
When  $\varpi$ is small in this case, we have:
\begin{equation}
\left\{
    \begin{aligned}
   \Omega^{\varpi_0} = \Omega^{GN} \left(1+ \frac{f_0^{\varpi_0}}{\varpi_0} \varpi + O[\varpi^2] \right) \text{ } \\
   R^{\varpi_0} = R^{GN} \left(1+  \frac{f_0^{\varpi_0}}{\varpi_0} \varpi + O[\varpi^2] \right). 
    \end{aligned}
    \right.
        \label{eq:par_zero}
\end{equation}
When $\varpi \gg \varpi_0$ in Eq.~\eqref{eq:s_w0}, we have: 
\begin{equation}
  \left\{
    \begin{aligned}
       \Omega^{\varpi_0} = \Omega^{GN} L_{\varpi_0} \left(1-\frac{f_{\infty}^{\varpi_0} \varpi_0^2}{\varpi^2} + O[\varpi^{-3}]\right) \text{ } \\
       R^{\varpi_0} = R^{GN} L_{\varpi_0} \left(1-\frac{f_{\infty}^{\varpi_0} \varpi_0^2}{\varpi^2} + O[\varpi^{-3}]\right).
     \end{aligned}
  \right.
 \label{eq:par_infinity}
\end{equation}
The HL approach can be included in this regime and from this equation we can understand why the HL solution overestimates $\Omega$ and $R$ with respect to GN, since it is given by the asymptotes:
\begin{equation}
  \left\{
    \begin{aligned}
       \Omega^{HL} = \Omega^{GN} L_{\varpi_0} \text{ } \\
       R^{HL} = R^{GN} L_{\varpi_0}.
     \end{aligned}
  \right.
\end{equation}
Moreover, it is possible to merge both behaviours of Eq.~(\ref{eq:par_zero}) and (\ref{eq:par_infinity}) into a single a function that goes continuously from one solution to the other:
\begin{equation}
  \left\{
    \begin{aligned}
       \Omega^{\varpi_0}(\varpi) = \frac{\Omega^{GN}\varpi_0^2+\Omega^{HL}\varpi^2}{\varpi_0^2+\varpi^2} - \frac{  \varpi \varpi_0^2 }{\frac{ \varpi^3}{\Omega^{HL} f_{\infty}^{\varpi_0}}  - \frac{\varpi_0^3}{\Omega^{GN} f_0^{\varpi_0}} } \text{ } \\
       R^{\varpi_0}(\varpi) = \frac{R^{GN}\varpi_0^2+R^{HL}\varpi^2}{\varpi_0^2+\varpi^2} - \frac{  \varpi \varpi_0^2 }{\frac{ \varpi^3}{R^{HL} f_{\infty}^{\varpi_0}}  - \frac{\varpi_0^3}{R^{GN} f_0^{\varpi_0}} }.
     \end{aligned}
  \right.
\end{equation}

\subsection{Samplings along the real axis}
%
Lets start by defining samplings, for now still with two frequencies, but that represent samplings along a line either parallel P or tilted, with a positive $A^+$ or negative $A^-$ angle, with respect to the real axis (RA).

\begin{equation}
   s^{RA}: \left\{
    \begin{aligned}
    z_1 = 0  \\
    z_2 = \omega_0
    \end{aligned}
    \right.
\end{equation}

\begin{equation}
   s^{P}: \left\{
    \begin{aligned}
    z_1 = i \varpi  \\
    z_2 = \omega_0+ i \varpi
    \end{aligned}
    \right.
\end{equation}

\begin{equation}
   s^{A+}: \left\{
    \begin{aligned}
    z_1 = 0  \\
    z_2 = \omega_0+ i \varpi
    \end{aligned}
    \right.
\end{equation}

\begin{equation}
   s^{A-}: \left\{
    \begin{aligned}
    z_1 = i \varpi \text{ } \\
    z_2 = \omega_0.
    \end{aligned}
    \right.
    \label{eq:s^A-}
\end{equation}

When $\varpi$ is small, we have:

\begin{equation}
\left\{
    \begin{aligned}
   \Omega^{P} =\Omega^{A+} = \Omega^{RA} \left(1- i \frac{f_0^{\omega_0}}{\omega_0} \varpi + O[\varpi^2] \right) \text{ } \\
   R^{P} = R^{A+} = R^{RA} \left(1-i  \frac{f_0^{\omega_0}}{\omega_0} \varpi + O[\varpi^2] \right), 
    \end{aligned}
    \right.
    \label{eq:samp_PA->0}
\end{equation}
which shows that a parallel sampling close to the real axis affects the parameters by a quantity that is proportional to the shift $\varpi$.

When $\varpi$ is large, we have:

\begin{equation}
  \left\{
    \begin{aligned}
       \Omega^{A+} = \Omega^{HL} \left(1-\frac{f_{\infty}^{\omega_0} \omega_0^2}{\varpi^2} + O[\varpi^{-3}]\right) \text{ } \\
       R^{A+} = R^{HL} \left(1-\frac{f_{\infty}^{\omega_0} \omega_0^2}{\varpi^2} + O[\varpi^{-3}]\right).
     \end{aligned}
    \label{eq:samp_A->I}
  \right.
\end{equation}


\subsection{Form of the $f$-factors  
\label{section:1p-factors}}

In order to get a general picture from the coefficients of the series expansions presented in the previous sections, we write them in a compact way by defining simply 
recognizable averaged quantities that may be weighted by particular elements. We use a notation, $\overline{\Omega^e}_{w}$,
in which $\Omega$ represent the average quantity, the position of a pole in this case, while $e$ is an exponent and $w$ the weighting function:

\begin{equation}
 f_\infty^{\omega_0} = \frac{1} {2 \omega_0^2} \frac{(\Omega_1^2 - \Omega_2^2)^2} {\overline{\Omega}_{R} \overline{\Omega}_{R^{-1}}} 
 \label{eq:f_in^0}
\end{equation}

\begin{equation}
 f_\infty^{\varpi_0} = \frac{1} {2 \varpi_0^2} \frac{(\Omega_1^2 - \Omega_2^2)^2} {\overline{\Omega}_{R} \overline{\Omega}_{R^{-1}}} 
 \label{eq:f_in^i0}
\end{equation}

\begin{equation}
 f_1^{0} = \frac{(\Omega_1^2 - \Omega_2^2)^2} {2 \Omega_1^2 \Omega_2^2} \frac{\varpi_0^2} {\overline{\Omega^2}_{x^{-1}}(i \varpi_0)} 
 \label{eq:f_1^0}
\end{equation}

\begin{equation}
 f_0^{\varpi_0} = \frac{(\Omega_1^2 - \Omega_2^2)^2} {(\varpi_0^2 + \Omega_1^2) (\varpi_0^2 + \Omega_2^2)} \frac{\varpi_0^2} {\overline{\Omega^2}_{x^{-1}}(i \varpi_0)} 
 \label{eq:f_0^omi0}
\end{equation}

\begin{equation}
 f_0^{\omega_0} = \frac{(\Omega_1^2 - \Omega_2^2)^2} {(\omega_0^2 - \Omega_1^2) (\omega_0^2 - \Omega_2^2)} \frac{\omega_0^2} {\overline{\Omega^2}_{x^{-1}}(\omega_0)} 
 \label{eq:f_0^om0}
\end{equation}

\begin{equation}
 L_{\varpi_0} = \frac{ \sqrt{ \overline{\Omega^{-2}}_{x}(i \varpi_0) }} {\overline{\Omega^{-1}}_{R}}
\end{equation}

\begin{equation}
   \left\{
    \begin{aligned}
     R_{HL} = R_1+R_2  \\
     \Omega_{HL} = \frac{1}{\overline{\Omega^{-1}}_{R}}
    \end{aligned}
    \right.
\end{equation}

\begin{equation}
 c_1^{0} = - \overline{x(\varpi_0)/x(i \varpi_0)}_{R/\Omega},
 \label{eq:c_1^0}
\end{equation}

where:

\begin{equation}
 \overline{\Omega}_{R} \equiv \frac{R_1 \Omega_1 + R_2 \Omega_2} {R_1+R_2}
\end{equation}

\begin{equation}
 \overline{\Omega}_{R^{-1}} \equiv ( \frac{\Omega_1} {R_1} + \frac{\Omega_2} {R_2} ) (R_1+R_2)
\end{equation}

\begin{equation}
 \overline{\Omega^{-1}}_{R} \equiv \frac{R_1 \Omega_1^{-1} + R_2 \Omega_2^{-1}} {R_1+R_2}
\end{equation}

\begin{equation}
 \overline{\Omega^{-2}}_{x}(i \varpi_0) \equiv \frac{\Omega_1^{-2} x_1(i \varpi_0) + \Omega_2^{-2} x_2 (i \varpi_0)}  {y(i \varpi_0)}
\end{equation}

\begin{equation}
 \overline{\Omega^2}_{x^{-1}}(\omega_0) \equiv  y(\omega_0) [\Omega_1^2 x_1^{-1}(\omega_0)+\Omega_2^2 x_2^{-1}(\omega_0)]
\end{equation}

\begin{equation}
 \overline{x(\varpi_0)/x(i \varpi_0)}_{R/\Omega}  \equiv \frac{\frac{R_1}{\Omega_1} \frac{x_1(\varpi_0)}{x_1(i \varpi_0)}+ \frac{R_2}{\Omega_2} \frac{x_2(\varpi_0)}{x_2(i \varpi_0)}} {\frac{R_1} {\Omega_1}+\frac{R_2} {\Omega_2}}
 \label{eq:c_1^0mx}
\end{equation}
